\documentstyle[12pt,equations,psfig,cite]{article}
\newenvironment{comment}[1]{}{}
\newcommand{\nn}{\nonumber}

\newcommand{\smallz}{{\scriptscriptstyle Z}} 
\newcommand{\smallw}{{\scriptscriptstyle W}} %
\newcommand{\smallh}{{\scriptscriptstyle H}} %
\newcommand{\smallv}{{\scriptscriptstyle V}}
\newcommand{\shat}{ {\hat s} }
\newcommand{\chat}{ {\hat c} }

\newcommand{\ecur}{ {\hat e} }
\newcommand{\as}{ { \alpha}_s }
\newcommand{\mz}{M_\smallz}
\newcommand{\mw}{M_\smallw}
\newcommand{\mh}{M_\smallh}
\newcommand{\mt}{M_t}
\newcommand{\mut}{\mu_t}

\newcommand{\seff}{\mbox{$\sin^2 \theta_{{ eff}}^{lept}$} }
\newcommand{\kf}{\mbox{$\hat{k}_f$}} 
\newcommand{\gc}{\mbox{$\hat{g}$}}

\newcommand{\scur}{\mbox{$\hat{s}^2$}}
\newcommand{\sincur}{\mbox{$\sin^{2}\!\hat{\theta}_{\scriptscriptstyle W} 
                           (\mz)$}}
\newcommand{\ccur}{\mbox{$\hat{c}^2$}}

\newcommand{\dr}{\mbox{$ \Delta r$}}

\newcommand{\drcar}{\mbox{$\Delta \hat{r}$}}
\newcommand{\drhoc}{\mbox{$\Delta \hat{\rho}$}}
\newcommand{\azz}{\mbox{$ A_{\smallz \smallz} $}}
\newcommand{\aww}{\mbox{$ A_{\smallw \smallw} $}}

\newcommand{\gd}{\mbox{$ O(g^4 \mt^2/ \mw^2)  $}}
\newcommand{\amtd}{\mbox{$ O(g^4 \mt^2 /\mw^2) $\ }}

\newcommand{\gq}{\mbox{$ O(g^4 \mt^4/\mw^4) $}}

\newcommand{\ew}{electroweak~}

\newcommand{\eps}{\epsilon}
\newcommand{\msbar}{\overline{\rm MS}}
\newcommand{\gsim}{\;\rlap{\lower 3.5 pt \hbox{$\mathchar \sim$}} \raise 1pt
 \hbox {$>$}\;}
\newcommand{\lsim}{\;\rlap{\lower 3.5 pt \hbox{$\mathchar \sim$}} \raise 1pt
 \hbox {$<$}\;}
\def\lequiv{\raise 0.4ex \hbox{$<$} \kern -0.8 em \lower 0.62 ex \hbox{$\sim$}}
\def\gequiv{\raise 0.4ex \hbox{$>$} \kern -0.7 em \lower 0.62 ex \hbox{$\sim$}}
\newcommand{\equ}[1]{Eq.\,(\ref{#1})}
\newcommand{\eqs}[1]{Eqs.\,(\ref{#1})}
\newcommand{\Eqs}[2]{Eqs.\,(\ref{#1}) and (\ref{#2})}
\newcommand{\efe}[1]{Ref.\cite{#1}}
\newcommand{\efs}[2]{Refs.\cite{#1,#2}}
\newcommand{\be}{\begin{equation}}
\newcommand{\ee}{\end{equation}}
\newcommand{\een}{\end{subequations}}
\newcommand{\ben}{\begin{subequations}}
\newcommand{\beq}{\begin{eqalignno}}
\newcommand{\eeq}{\end{eqalignno}}
\newcommand{\bea}{\begin{eqnarray}}
\newcommand{\eea}{\end{eqnarray}}
\def\pl#1#2#3{{\it Phys. Lett. }{\bf B#1~}(19#2)~#3}
\def\np#1#2#3{{\em Nucl. Phys. }{\bf B#1~}(19#2)~#3}
\def\prl#1#2#3{{\it Phys. Rev. Lett. }{\bf #1~}(19#2)~#3}
\def\pr#1#2#3{{\em Phys. Rev. }{\bf D#1~}(19#2)~#3}
\def\zp#1#2#3{{\it Z. Phys. }{\bf C#1~}(19#2)~#3}

\newenvironment{appendletterA}
 {
  \typeout{ Starting Appendix \thesection }
  \setcounter{section}{0}
  \setcounter{equation}{0}
  
 }{
  \typeout{Appendix done}
 }
\newenvironment{appendletterB}
 {
  \typeout{ Starting Appendix \thesection }
  \setcounter{equation}{0}
  
 }{
  \typeout{Appendix done}
 }

\renewcommand{\thefootnote}{\fnsymbol{footnote} }
\begin{document}
\begin{titlepage}

\begin{flushright}
        \small
        DFPD--99/TH/19\\
        TUM--HEP--333/98

\end{flushright}
\vspace{1.2cm} 
\begin{center}
{\Large\bf 
Two-loop heavy top corrections\\
\vspace{2mm}
to the $Z^0$ boson partial widths}\\ 
\vspace{2cm}
{\sc Giuseppe Degrassi$^a$ and 
 Paolo Gambino$^b$}\\
\vspace{.4cm}
{\em $^a$ Dipartimento di Fisica, Universit\`a  di Padova and\\ 
INFN, Sez.
di Padova, via F.~Marzolo, 8, I-35131, 
Padova, Italy. }\\
\vspace{.3cm}
 {\em $^b$ Physik Dept.,
Technische Universit\"at M\"unchen,\\
James-Franck-Str., D-85748 Garching, Germany}
\end{center}
\vspace{2cm} 
\begin{center}
{\bf Abstract}
\end{center}
\vspace{0.2cm}
We present the evaluation of  the two-loop $\gd$ effects
in  the partial widths of the $Z^0$ boson in the $\msbar$ scheme
and in two different implementations of the on-shell scheme.
We observe a clear reduction of the scheme dependence of the predictions.
The renormalization procedure 
and the Heavy Top Expansion employed in the \gd\ calculations
are illustrated in some detail and  intermediate results are provided. 
We discuss the implication of our results 
on the constraints for the Higgs mass making use of simple
interpolating formulas. 
We find that precision data give $\mh < 285$ GeV at 95\% C.L. 
taking into account the theory uncertainty. Including also the information 
from  direct search experiments we obtain a 95\% upper bound $\mh < 345$ GeV. 
\vfill\nopagebreak

\end{titlepage}
\setcounter{footnote}{0}
\renewcommand{\thefootnote}{\arabic{footnote}}

\section{Introduction}
The impressive  amount of data collected at LEP, SLC,  and the Tevatron
allows for a very stringent test of the Standard Model (SM), at the level of
the per mille accuracy. This experimental situation has required a significant
effort from the theory side in order to provide predictions
for the electroweak observables with comparable  precision. 
We recall that such a level of accuracy tests the quantum structure
of the theory, thereby providing information about physics at mass scales
that are not directly accessible at our experimental facilities. 

After the discovery of the top quark at Tevatron the only  mass scale 
still unknown in the SM  is the one related to the Higgs boson. 
As the sensitivity of the various 
precision observables to the Higgs mass, $\mh$, is approximately 
logarithmic, the extraction of the relevant
information is  more difficult and delicate than in the case of the top 
and requires a careful consideration of higher order effects.
In the recent past it was recognized \cite{YB,dffgv} that
in order to obtain accurate constraints on $\mh$ 
from global fits to the experimental data, the theoretical predictions
of the observables most sensitive to $\mh$ should also include
two-loop top effects enhanced by factors $(\mt^2/\mw^2)^n\:(n=1,2)$. 
This observation called for the study of the \gd\ corrections
to the $\mw-\mz$ interdependence,  presented in \efe{dgv}, 
and  to the effective mixing parameter \seff \cite{dgs}. 
Moreover, in \efe{dgs} the analysis of the $\gd$  contributions
to $\mw$  and $\seff$ was carried out    in the 
$\msbar$ formulation of \efe{dfs} as well as in two  different implementations
of the on-shell (OS) scheme \cite{si80}. It was shown that 
the inclusion of \gd\ effects
sharply reduces the scheme and residual scale dependence of the
theoretical predictions for $\mw$  and $\seff$, and suggests a comparable
reduction of the overall theoretical uncertainty \cite{dgps,barcelona}.

The aim of this paper is to go further in  the study of the \gd\
corrections. Our purpose is actually manifold:
\begin{itemize}
\item First,  we   present the result of the
calculation of the partial widths of the $Z^0$,
$\Gamma_f$, for $f\neq b$, at $\amtd$ in both the $\msbar$ and OS
schemes. As the case of the $b$ quark involves  
a completely new class of vertex diagrams and would require a significant 
dedicated effort, it will not be considered here. 
We also study the scheme dependence of the results
for the leptonic width, $\Gamma_\ell$, 
finding it significantly  reduced by the inclusion of the new contributions.

\item We take the opportunity of the new result on the partial widths to
describe   some  important aspects common to all \gd\  calculations 
which were not 
discussed in the preceding publications, in particular concerning the 
renormalization procedure  and the Heavy Top  Expansion (HTE) technique used 
in the calculations. 
We also provide some intermediate results and derivations in appendix A.

\item Finally, we discuss some of the consequences of our results, in view
of the present experimental situation. 
Following \efe{dgps}, we provide elementary interpolating formulas in 
different schemes for $\Gamma_\ell$. For completeness, we also report and 
update the analogous formulae for $\mw$ and $\seff$.
\end{itemize}

A few  topics relative to the \gd\ calculations
were already covered in some detail in preceding 
related publications, and we will not dwell on them further. 
For instance, the impact of the new calculations on the indirect
determination of $\mh$ is quite important, and this has been extensively 
discussed in Refs.\cite{dgs,dgps,peppe,barcelona}.
Although we do  not repeat this discussion here, 
 we will use  the interpolating formulae for $\mw$, $\seff$, and 
$\Gamma_\ell$,
together with the latest experimental data, to obtain two new estimates
for $\mh$. One is based
 on information coming only from precision measurements, while the other 
includes also the results of direct search experiments, using the method
developed in \efe{higgsbound}.

The paper is organized as follows. In the next section we discuss some
issues concerning the renormalization procedure we have followed. Section
\ref{hte} deals with the technical matter of the HTE. In 
section~\ref{widths} we present the analysis of the corrections to the $Z$ 
partial widths to \gd. Section 5 contains the interpolating
formulae and our estimates for $\mh$. Finally, we present a brief summary.
We also include two appendices. In appendix A we provide  
the two-loop $W$ and $Z^0$ self-energies to \gd,
which may be useful in different contexts. 
Appendix B contains  instead  the explicit expressions of the \gd\  
corrections to  the various $Z$ partial width form factors.

\section{Renormalization}
\label{renorm}
In this section we  discuss some  aspects of the \amtd
renormalization for  processes involving four light fermions as
external states. The main difference between the one and two-loop 
cases is that in the latter  
we also have to consider the renormalization of   the unphysical sector.
In the following, we first discuss  the renormalization of 
the unphysical sector in a general
way. We  then explain why the use of OS masses for the vector bosons, 
rather than $\msbar$ mass parameters, 
is particularly useful in our calculations
 and illustrate different options for the 
renormalization of their unphysical counterparts. 

As is well known, there are constraints that link the longitudinal component 
 of the vector bosons, the corresponding pseudo-Goldstone bosons, and the 
Faddeev-Popov (FP) ghosts.
In higher order calculations, 
it is convenient to choose a renormalization procedure that 
automatically  respects the Slavnov-Taylor Identities (STI)
which are induced by the local gauge invariance of the original Lagrangian
before spontaneous symmetry breaking.
According to the organization of the calculation,
it is possible to use different procedures that respect the STI's
and are particularly convenient in order, for instance, 
to minimize the number of diagrams  to be
considered. Of course, physical amplitudes are independent of the 
chosen procedure, and this can be used as an
additional check of the calculation.
For the problem at hand, the discussion can be kept at the one-loop level, 
and we can limit ourselves to  contributions proportional to powers 
of $\mt$. 
In the following, we discuss only the $W$ sector; the case of the $Z^0$ 
sector can be treated in a completely analogous manner, because  
the $\gamma-Z^0$ mixing does not contribute   at the order of our calculation. 

If we split the unrenormalized $W$ polarization tensor  into 
transverse and longitudinal parts
\be
\Pi^{\mu\nu}_\smallw(q)= \left(g^{\mu\nu} -\frac{q^\mu q^\nu}{q^2}\right) 
A_{\smallw\smallw}(q^2)
+ \frac{q^\mu q^\nu}{q^2} \Pi^{{\scriptscriptstyle L}}_{\smallw\smallw}(q^2),
\ee
and we denote by $\Pi_{\smallw\phi}$ 
the proper two-point function for the  mixing between the $W$ and its
pseudo-Goldstone boson, $\phi$, and  by $\Pi_{\phi\phi}$ the self-energy of the
latter, 
we obtain the following STI's in a general $R_\xi$ gauge 
(see for ex. \cite{aoki:82,boehm:86})
\bea
&&\xi \left[\Pi^{{\scriptscriptstyle L}}_{\smallw\smallw}(q^2) +
\mz \Pi_{\smallw\phi}(q^2)\right] +G(q^2)=0
\nonumber\\&&
q^2\left[ \Pi^{{\scriptscriptstyle L}}_{\smallw\smallw}(q^2) + 2 \mw 
\Pi_{\smallw\phi}
(q^2)\right] + \mw^2 \,\Pi_{\phi\phi}(q^2) + \mw^2 \,T=0\, ,
\label{ident}
\eea
where $T$ represents the tadpole  contribution and $G(q^2)$ is a term 
involving charged FP  ghosts as external fields.
Among other things, $G(q^2)$ contains 
 the  two-point function of the charged  ghosts;  as the ghost do not
 couple directly to fermions, at the one-loop level $G(q^2)$ is completely 
independent of the top quark. 

We see that the first two terms of the second STI in \eqs{ident}
vanish  at $q^2=0$, and
that therefore $\Pi_{\phi\phi}(0)$ is equal to  the tadpole contribution. 
This uncovers 
the connection between  the renormalization of the Goldstone boson mass 
and  the one of the tadpole. 
\begin{comment}{As anticipated, the counterterm contributions to the various 
terms in \equ{ident} must also respect the STI's.}\end{comment}
In our calculation we employ
the usual tadpole renormalization that minimizes the effective 
potential and consists in removing all tadpole graphs. 
This choice implies the 
subtraction of $\Pi_{\phi\phi}(0)$ from the two-point function of the
pseudo-Goldstone boson  \cite{taylor,zucchini}.

Concerning  the longitudinal component of the two-point function of the $W$, 
we find it convenient to renormalize it in the same way as the
transverse part, the one which  is related to the mass counterterm.
Therefore, the choice of the
renormalized mass parameters employed in our analysis is going to affect
the structure of the finite part of the counterterm in the unphysical
sector. 
In practice, the choice is between the  $\msbar$ and OS mass definitions.
 For the other two-point functions appearing in \equ{ident}
different options are possible, which all
respect the STI's, and are equivalent at the level of physical amplitudes.
They correspond to different ways of renormalizing the gauge-fixing parameters.

A first and very natural option ({\it heavy mass} procedure)
consists in subtracting from  each term in \eqs{ident}
the contributions proportional to powers of $\mt$. 
In case  an $\msbar$ prescription for the vector boson masses is adopted,
the subtraction applies only to the divergent part.
This procedure is equivalent to a Taylor expansion in the external momentum 
and  obviously respects the identities, as the 
unrenormalized self-energies do. We recall that 
$ \Pi^{{\scriptscriptstyle L}}_{\smallw\smallw}(0)=A_{\smallw\smallw}(0)
\approx {\rm Re}\,A_{\smallw\smallw}(\mw^2)$
at the order we are interested in, so that the OS mass counterterm 
corresponds to the first term of the Taylor expansion.
There is  a counterterm 
for the $W-\phi$ transition as well as for the $\phi^\pm$ two-point function,
but the FP ghost self-energy needs not be renormalized, as it is independent  
of the top quark. 
In order to satisfy  the second of \eqs{ident} at any $q^2$,
a wave function renormalization  $Z_{\phi}$ for 
the $\phi^\pm$ field is needed. On the other hand, $\phi^\pm$ 
appear only inside the loops in our calculations and the use of 
$Z_{\phi}$ is indeed not necessary. 
 Therefore, we simply demand that
the second of \eqs{ident} be satisfied at $q^2=\mw^2$.

A second possibility consists in assigning  no counterterm 
to the $W-\phi$ transition; this decreases the number of counterterms and
simplifies  their implementation. In fact, in the 't~Hooft-Feynman gauge 
it corresponds to renormalizing the masses of the 
vector boson and of the associated scalar boson and ghost in the same way
({\it bare gauge fixing}), apart from the  supplementary subtraction at 
$q^2=0$, corresponding to the tadpole contribution, that we have discussed 
above. It therefore amounts to fix $\delta M^2_\phi=\delta \mw^2 + 
T$ and $\delta M^2_c=\delta \mw^2$ ($c^\pm$ are the FP ghosts)
and  is the closest to the naive parameter renormalization.
Also this choice  verifies both \eqs{ident} at $q^2=\mw^2$, leaving 
room for a further arbitrary  wave-function renormalization, which again can 
be  avoided altogether.

We have explicitly verified that these two renormalization options 
({\it heavy mass} and {\it bare gauge fixing})
 are equivalent at the level of physical amplitudes in the cases at hand.
Finally, we should mention that 
another  possibility is the one presented in \cite{rosstaylor},
which corresponds to the use of {\it renormalized gauge-fixing}.

For what concerns the three-point functions,
 we just recall  that the renormalization of those couplings that contain   
masses is fixed by the Ward identities that link the 
Yukawa and bosonic coupling  counterterm 
to the gauge coupling and  mass renormalization. 

\begin{figure}[t]
\centerline{
\psfig{figure=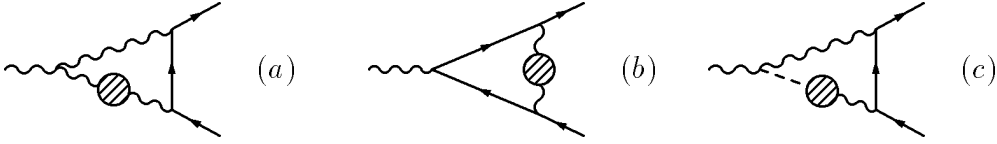,rheight=0.73in}  }
\caption{\sf Two-loop vertex diagrams containing the top. Wavy (dashed) 
lines represent vector (scalar) bosons. }
\label{vert}
\end{figure}

From a technical point of view, 
the choice of OS masses as  renormalized parameters, in particular for the 
$W$ and $Z^0$ vector bosons,  has some important 
advantages in the calculations considered here.  
The processes we are interested in always involve 
light external fermions and are mediated  at lowest level 
(to excellent approximation)  by gauge bosons only.
In these cases, the two-loop box and vertex diagrams containing the top 
quark 
consist of a top-bottom insertion on a  bosonic line belonging to a one-loop
vertex (Fig.\ref{vert}a,b) 
or box\footnote{A top loop vertex insertion on a trilinear coupling
does not provide $\mt^2$ terms by dimensional reasons.}, plus vertices that
involve a $W-\phi$ transition, like the one shown in Fig.\ref{vert}c.
Superficially, all these diagrams  are  either convergent
or  logarithmically  divergent. However, they 
contain a quadratic subdivergence associated with the  top-bottom loop 
insertions. 
If we employ the OS mass definition for the mass of the vector bosons,
all the boxes and vertices involving insertions on vector boson lines
can be  neglected in the calculation because the associated mass counterterm
graphs remove completely their $\mt^2$ dependence.
Of course, this would not happen in the case of a pure $\msbar$ subtraction,
as the remaining finite part would still be proportional to $\mt^2$. 

The use of OS vector boson masses 
 therefore absorbs a whole class of \amtd \ diagrams
and we have decided to   present always our results in terms of
OS masses for all the physical particles, 
also when  we work in the $\msbar$ scheme.
Indeed, this choice has some clear advantages 
even in the case the gauge couplings are renormalized in  the $\msbar$ 
scheme \cite{Si89,dfs}: (i) it 
employs as mass parameters the physically measured masses without introducing 
large radiative corrections in the coupling counterterms;
(ii) it absorbs the large radiative corrections
that appear in the one-loop relation between OS and $\msbar$ masses.  
Incidentally, one should also  note that the conventional OS mass definition, 
identified by the counterterm ${\rm Re}\,A_{\smallw\smallw}(\mw^2)$,
does not present any gauge ambiguity at the order of our 
calculations; as stressed in \cite{polemass}, this is not the case at $O(g^4)$ 
and beyond. 

The {\it heavy mass} procedure that assigns a counterterm  to the
$W-\phi$ transition has an additional advantage: it
allows us to neglect systematically 
all two-loop vertices and boxes in our calculations, including the diagrams 
of Fig.1c. In this case
the only two-loop diagrams that we need to consider are two-point functions
of the vector bosons:  this is the approach followed in the present paper. 
If instead we work with {\it bare gauge fixing}, it should be apparent
 that only diagrams of the kind in 
Fig.\ref{vert}(a,b) will be removed at \amtd\ after renormalization. 
The diagram of Fig.\ref{vert}(c) has to be explicitly calculated. However, as
could be expected, its evaluation  at \amtd\ involves only 
self-energy integrals
and therefore  no additional difficulty is introduced. This approach was
followed in \efs{dgv}{dffgv} and the detailed evaluation of the diagram of 
Fig.\ref{vert}(c) was discussed in \efe{zako}\footnote{Independently of 
the renormalization of the unphysical sector,
vertex diagrams of the kind shown in Fig.\ref{vert}(c) can be excluded 
by the choice of a non-linear gauge fixing; 
for example the $\gamma W^\pm\phi^\mp$ coupling, which is
characteristic of the $R_\xi$ gauge, can be  avoided, so that the photon 
separately couples to $W$ and charged Goldstone bosons\cite{fujikawa}.
Consequent 
 modifications \cite{barroso} in the Faddeev-Popov sector of the SM do
not affect the calculation, as ghosts do not couple to fermions.
In the case of the photon vertex,  
also the background field gauge can be used to the same effect. 
Alternatively, it is possible to fix the gauge in such a way that the
$Z^0 W^\pm\phi^\mp$ vertex is avoided \cite{fawzi}.}.

\section{Heavy top expansion of two-loop diagrams}
\label{hte}
In the preceding section we have seen that in the calculations at hand 
the only two-loop diagrams that need to be considered are two-point
functions of the vector bosons, including the $\gamma-Z^0$ mixing for the
calculation of the effective sine and the photon self-energy at vanishing
momentum for the renormalization of the electric charge \cite{si80}.
In this section we explain how we have evaluated
these self-energy diagrams.
We start motivating the choice of the next-to-leading order (NLO)
HTE instead of a numerical computation, proceed to consider the classes of 
diagrams that need to be computed, and conclude discussing
 the quality of the approximation. 

There are arguments \cite{scharf} that 
two-loop self-energy integrals with arbitrary masses and momentum transfer 
cannot be  expressed
in terms of known elementary functions like polylogarithms.
Although an analytic solution seems in the general case impossible,  
there exist several methods that permit an efficient numerical evaluation
\cite{numerical}. 
On the other hand, compact analytic results are available in a number of 
special cases, which apply to a variety of physical situations like small
and large momentum expansions. 
The clear advantage of the 
analytic evaluation in a situation where at least one
of the parameters, namely $\mh$, is largely  unconstrained, 
is that the results can be
readily implemented in fitting routines.
Furthermore, if one chooses to perform a numerical calculation 
at the two-loop level in the \ew theory without 
any approximation, several challenging problems arise,
 like the  very  large number of diagrams, 
the implementation of the  complete two-loop renormalization of the SM, etc.
Therefore, following the conclusions of \cite{YB,dffgv}, 
our strategy has been to use a heavy mass expansion up to \gd.

\begin{figure}[t]
\centerline{
\psfig{figure=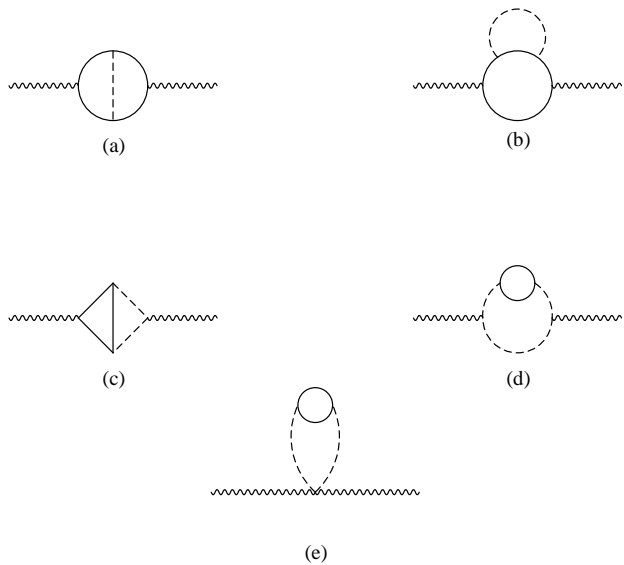,height=3.in,rheight=2.8in}  }
\caption{\sf Two-loop topologies for vector boson self-energies. 
Dashed (solid) lines represent virtual bosons (fermions). }
\label{lead}
\end{figure}

We have performed the calculations in the 't~Hooft-Feynman gauge, where
$M_{\phi}=M_{c}=\mw$. This choice proves to be convenient also 
in the evaluation of the integrals, as it limits
the number of different mass
scales occurring in each diagram. Concerning the bottom mass, we have set it
to zero from the start.  We have neglected all flavor violation. 
All two-loop self-energy diagrams 
containing the top have to be considered. The topologies  
are displayed in Fig.\ref{lead}.
The  subset of diagrams containing only scalar bosons in addition to the top
and the bottom quarks is responsible for the leading \gq\ correction
\cite{barb}.

We distinguish between three classes of diagrams: 
(i) the ones for which a naive Taylor expansion in $q^2$ can be used because no
light mass threshold is involved;
(ii) those containing at least one bottom quark threshold, 
which we regard  as zero mass threshold;
(iii) those with at least one physical threshold of the order of the 
$W$ mass (e.g. it can be $4\mw^2$, $\mw^2$, $(\mw+\mz)^2$).
Clearly, the two last classes of diagrams overlap  (see for ex. 
the diagram in Fig. \ref{lead}(c) where the fermion loop consists of
 two bottom and one top lines and the bosonic lines represent two $W$ bosons).
With respect to this classification of diagrams, the fact that the Higgs boson
mass is {\it a priori} arbitrary implies that we have 
to consider two scenarios:
(a) the case in which $\mh$ is light compared to the top mass, namely
$\mh\approx\mw$ and therefore $\mh\ll\mt$; (b) 
the case in which $\mh$ is considered heavy compared
to $\mw$, and the heavy mass expansion is implemented leaving the ratio
between $\mh$ and $\mt$ finite and arbitrary.
For most  diagrams containing the Higgs, the HTE provides different expressions
in the two cases. This explains why in app.\,A the expressions for the
$W$ and $Z^0$ self-energies are given in the two regimes. 

In the case of class (i) diagrams we adopt a Taylor expansion  
first in the small external momentum and then in the light
masses keeping only terms proportional to powers of $\mt$. 
The Taylor expansion of the two-loop integrals
is viable for values of $q^2$ up to the first physical threshold. Since
in our calculations $q^2=M^2_{\smallw,\smallz}$ or 0, it follows that
$q^2\ll\mt^2$ and this procedure is justified for class (i).
\begin{figure}[ht]
\centerline{
\psfig{figure=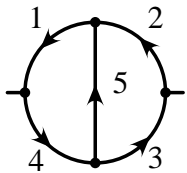,height=1.in}  }
\caption{\sf Two-loop master self-energy diagram. }
\label{master}
\end{figure}
Typical diagrams belonging to this
class are those of Fig.\ref{lead}(a) where the fermion in the loop is a top
quark. Referring to the two-loop master diagram of Fig.\ref{master}, 
any diagram containing a heavy particle (top and possibly  Higgs boson)
in place 1 {\it and} 2 can be safely included in this class.
In practice, the $q^2$ expansion
 is implemented  using ($k$ and $m$   generically indicate one 
momentum of integration and one internal  mass) 
\be
\frac1{(k+q)^2 -m^2}\to \sum_{n=0}^{\infty}
(-1)^n \,\frac{(2k\cdot q +q^2)^n}{(k^2-m^2)^{n+1}}
\label{taylor}
\ee
under the $n$-dimensional  integral. For arbitrary internal masses, 
the  two-loop integrals  at zero momentum transfer
can be expressed as a combination of logarithms and dilogarithms \cite{DT}.
Therefore, we reduce the integrand to scalar vacuum integrals 
and for the evaluation of the 
integrals with higher powers of the propagators 
we follow the iterative method of first paper in \cite{DT}. 

For the diagrams of class (ii)  the  Taylor expansion  is bound
to fail because it cannot describe the physical threshold at $q^2=0$.
The typical diagram has bottom at place 1 and 4 and top at place 2 and
3 in Fig.\,\ref{master}. The simple replacement of \equ{taylor} 
when $m=0$ leads to 
increasingly severe infrared divergences  for increasing $n$.
Heavy mass or small momentum
expansions in these cases are therefore  asymptotic expansions, about which a
rich technical literature exists \cite{expa}. 
\begin{comment}{Before an expansion of the kind of \equ{taylor} 
can take place,  the sub-integrations responsible for the appearance of the 
the infrared divergences have to be subtracted.
From a theoretical point of view, this is equivalent to 
the construction of an off-shell effective theory  at some order in the 
heavy mass scale (in our case $\mt^2$). After computing the coefficients
of the composite operators, their  insertions into   one-loop 
integrals with  thresholds at $q^2=0$ ensures that the analytic structure 
of the amplitude is preserved.
An elegant algorithm \cite{expa} has been developed in dimensional
regularization which implements the effective theory at the diagrammatic level.
The algorithm has been applied to this class of diagrams in \cite{zerothres}.
}\end{comment}
An elegant algorithm  to implement the asymptotic expansion
in dimensional regularization has been developed in \efe{expa}
and applied to this class of diagrams in \cite{zerothres}.
The technique  we have used, however, is based on  the 
repeated application of the identity
\be
\frac1{(k_1+k_2)^2-M^2} = \frac1{k_1^2-M^2} -
 \frac{k_2^2 +2 k_1\cdot k_2}{[(k_1+k_2)^2-M^2](k_1^2-M^2)}
\label{idint}
\ee
where $M$ is a heavy mass and $k_{1,2}$ are the integration momenta.
The first term on the r.h.s.~of \equ{idint} leads to a disconnected 
integral (product of two one-loop integrals) that can be evaluated exactly,
i.e.~for arbitrary $q^2$, and contains  $\ln q^2$ terms that describe 
 the physical threshold.    The second term, instead, 
leads to  a two-loop integral with improved infrared convergence in the $k_2$
integration and improved ultraviolet convergence in the $k_1$ integration.
Therefore, if, for example,  the original integral is IR logarithmically
divergent in the $k_2$ integration when $q \to 0$, the corresponding 
two-loop integral 
associated with the second term in \equ{idint} evaluated at $q^2=0$ 
gives a finite result that differs from the complete result for 
arbitrary $q^2$ by terms $O(q^2 /M^2 \ln (q^2/M^2))$. In case of 
integrals more seriously IR divergent, an iterative 
use of \equ{idint}  gives 
an approximate evaluation,
up to terms suppressed by the heavy mass, through a combination of
integrals (a set of disconnected ones plus two-loop integrals 
evaluated at $q^2=0$) whose analytic expressions are known.
Thus, one can combine the expansion (\ref{taylor}) with the
repeated use of \equ{idint} to
obtain the asymptotic expansion of the original integral at the desired
order. Indeed,  at each step of the expansion  the more serious
IR divergent behavior introduced by (\ref{taylor}) can be balanced by the 
convergent IR factor from  the  second term of \equ{idint}, while
the part associated with the first term in \equ{idint} can be always evaluated
exactly. An explicit example of the application of this method 
has been provided in \cite{zako}.
We have checked that our results coincide 
with those of \efe{zerothres} in all the cases relevant to the present 
calculation.

The class (iii)  diagrams differs from those of class (ii) because the
threshold is not at $q^2=0$ and can in fact be well above the $q^2$ at which 
the integrals are evaluated. Nevertheless, the Taylor expansion cannot be
employed here 
because  we are interested in the HTE and not in the small momentum
expansion. Therefore, in order to extract all the terms enhanced by powers
of $\mt$, one has to use
the same procedure adopted for case (ii), reducing the original integral
to a sum of disconnected ones, to be evaluated exactly keeping
the light masses ($\mw,\,\mz$ and possibly $\mh$) and the $q^2$, plus two-loop 
connected integrals with  $q^2=0$ which have to expanded in powers of $\mt^2$.
The disconnected diagrams are responsible for the terms containing the
functions $B_0[q^2,m_1^2,m_2^2]$ in the two-loop self-energies  of app.\,A. 
Also in this case we have checked our results against the relevant
existing literature \cite{berends}.

A crucial question is now whether the approximation  based on the 
NLO HTE described above is reliable.
The second term $O(g^4 \mt^2/\mw^2)$ of the two-loop HTE seems to be
quite important wrt the first \cite{dgv}, 
so the convergence of the HTE may be legitimately questioned. 
An important point to take into account in this
respect is that this is true mainly for a light Higgs,
 where the approximation
of keeping $\mw=0$ and $\mh\neq 0$ manifestly fails. The result of  
\cite{barb}, which was based on such approximation, 
becomes meaningless in this regime,  and no hierarchy 
among the first and the second term of the asymptotic expansion should
 be expected. 
Moreover, we know from  \efe{dgs} that the size of the \gd\ contributions to
 $\seff$ and $\mw$ depends very strongly on the scheme adopted. 
This tells us that
reducible contributions (products of one-loop integrals) are 
very important there. Since we know that the HTE
works very  well at one-loop level where the leading quadratic
term is dominant,  the reducible contribution is well approximated by the 
first two terms of the HTE.
Concerning the irreducible contributions, 
the present two-loop calculations are  based on two-point
functions only. Unlike the case of three and four point-functions~\cite{box}, 
the HTE appears to work quite well for self-energies,
as has been demonstrated in the case of QCD corrections up to
three loops in the second paper of \efe{qcd3}.
Ultimately, however, the most convincing evidence on the quality of the NLO HTE
approximation comes {\it a posteriori} from the sharp reduction of the scheme
and scale dependence of the predictions for $\mw$, $\seff$, and, as will be
seen later, $\Gamma_\ell$. 

A conclusive test of our approximation can only come from a  comparison  
of the HTE  with a two-loop numerical calculation.
A first partial test of the HTE \cite{gsw} for the two-loop \ew corrections 
has  been obtained by comparing the results
of \cite{dgv,dgs} with the calculation of \efe{weiglein} by
Bauberger and Weiglein. They  calculated  the two-loop self-energies
which contain the Higgs boson together with fermions through a
direct numerical evaluation of Feynman diagrams, without any
heavy mass expansion.
Although this is only a subset of the diagrams involving the Higgs boson,
the result can be used as a partial test of the HTE employed here, because it
includes all the diagrams containing both top and Higgs.
The Higgs mass dependences of the predictions for $\mw $ and $\seff$ 
according to Refs.\cite{dgs} and 
\cite{weiglein} have been accurately compared in 
\cite{gsw,barcelona}. Very small deviations appear to be attributable
to the use of the HTE, of the order of 1 MeV and $10^{-5}$ for $\mw $ and
$\seff$, respectively. Such deviations 
are well within the range of theoretical errors estimated
from scheme and scale dependence \cite{dgs,barcelona}.

\section{The  $Z^0$  decay partial widths}
\label{widths}
This section contains the main result of the paper, namely the
$\gd$ corrections to the partial decay widths
$Z \rightarrow f \bar{f},\: \Gamma_{f}$, of the $Z^0$ boson. 
The only case which is 
not covered is the one of final $b$-quarks, because it involves 
specific \gd\  vertex corrections.
As in our previous calculations \cite{dgv,dgs}, we first derive the
result in the $\msbar$ framework of \efe{dfs} and then discuss
the  scheme dependence by considering the analogous contributions 
in two implementations of the OS scheme. 

We recall that the
formulation of \efe{dfs} employs $\msbar$ couplings evaluated at the scale 
$\mz$ while using OS masses for the 
physical particles. In particular, the coupling constants of the
SU(2) and U(1) groups are written in terms of 
$\ecur^2 = e^2/(1+ 2 \,\delta e/e)_{\msbar}$, the $\msbar$ electromagnetic 
coupling,
and $\sincur \equiv \shat^2$,  the $\msbar$ electroweak mixing 
parameter. Even though the top quark is heavier than the $Z^0$ mass scale, 
it is not decoupled in the definition of the $\msbar$ couplings,
 according to the convention of \efe{dfs}.
 The subscript $\msbar$ in the charge
renormalization counterterm $( 2 \,\delta e/e)$ 
indicates  the pole subtraction, 
a notation that will also be used henceforth. 
Although the t'Hooft scale $\mu$ is conventionally set equal to $\mz$, in
our calculations we leave it unspecified. This allows us 
 to study the scale dependence of the results. 
In this framework, $\mw$ and the renormalized parameter $\scur$ are 
calculated from  $G_\mu$,  $\alpha$,  and $\mz$ 
through the two relations ($\chat^2 = 1 - \scur$)
\be
 \frac{\mw^2}{ \mz^2 }= \frac{ \chat^2}{1-\frac{\hat{e}^2}{\hat{s}^2}\drhoc},
\ \ \ \ \ \ \ \ \ \ \ \ 
\scur  = \frac{\pi \alpha}{\sqrt{2} G_\mu\mw^2 (1- \Delta \hat{r}_\smallw)}.
\ee
The \gd\ contributions to the radiative corrections 
$\frac{\hat{e}^2}{\hat{s}^2}\drhoc$ and $\drcar_\smallw$
were discussed in \efs{dgv}{dgs}. In appendix B we report  the explicit 
expression for the \gd\ contribution  to $(2\, \delta e/e)_{\msbar}$.
We notice that there is a  technical advantage in using $\msbar$ 
couplings. Indeed, the associated one-loop
counterterms, $\delta^{(1)} \hat{e}$  and $\delta^{(1)}\scur$, are pure
poles that  do not contain any $\mt^2$ dependence and this
reduces the number of counterterm diagrams to be considered.  

The amplitude for the decay of a $Z^0$ boson into two fermions $f$ 
can be expressed in the $\msbar$ scheme as \cite{ds91}
\be
M^\mu_Z= -i \frac{\ecur }{\shat \chat} \sqrt{\hat{\eta}_f}
\,\langle f \bar f |\, \frac1{2}  J_3^\mu - \scur \,\kf (\mz^2)
\, J_\gamma^\mu \,|0\rangle,
\label{ampli}
\ee
where   $J_3^\mu$ is the $SU(2)$ neutral 
current and $J_\gamma^\mu$ is the photonic 
current\footnote{In \efe{ds91} the correction $\hat{\eta}_f$ is indicated as
$\bar\rho_{ff}$.}. 
The corresponding expression for the partial  width $ Z \rightarrow f \bar f$ 
is therefore
\be
\Gamma_{f}= N_c^f\, \frac{\ecur^2}{\scur \ccur} \frac{\mz}{96\pi}
|\hat{\eta}_f| \left\{ 1 - 4 \,I^3_f \,Q_f \,\scur {\rm Re}\hat{k}_f + 8 \,Q_f^2 \,
\hat{s}^4 |\hat{k}_f|^2 \right\} + O(m_f^2/\mz^2),
\label{gamma}
\ee
where $N_c^f$ is the color factor for the fermion $f$,
$I^3_f=\pm1$ its isospin quantum number, $Q_f$ its electric charge, and
the appropriate  QED and QCD corrections have not been
explicitly indicated.

The \gd\ corrections to $\kf(\mz^2)$ have been already considered in 
Refs.\,\cite{dgs,zako}.
In the $\msbar$ scheme they are universal (with the exception of the $b$
final state) because  they  can be entirely expressed 
in terms of  the $\gamma-Z$ self-energy, 
as a consequence of the discussion in  Sec.\ref{renorm}. 
The overall factor  $\hat{\eta}_f$ can be expressed as $
\hat{\eta}_f = 1 + (\ecur^2 /\scur ) (\Delta\hat{\eta}_f^{(1)} +
\Delta\hat{\eta}_f^{(2)})$,
where  the one-loop contribution is given by 
\be 
(\ecur^2 /\scur) \Delta\hat{\eta}_f^{(1)}
=  \left. {\rm Re} A_{\smallz\smallz}^{(1)'}(\mz^2)\right|_{\msbar} 
+\left.\frac{\hat{g}^2}{16\pi^2} {\bar V}_{ff}(\mz^2)\right|_{\msbar} .
\ee
The function  ${\bar V}_{ff}(\mz^2)$ is defined  in Eq.\,(12d) of \efe{ds91} 
and 
${\rm Re} A_{\smallz\smallz}^{'(1)}(\mz^2)$ is given in the 
appendix of the same paper. 

In analogy to the case of $\kf$, 
the $\msbar$ two-loop \gd\ contribution  $\Delta\hat{\eta}_f^{(2)}$ 
is the same for all 
light fermions with the exception of the bottom. Indeed, we have seen  in 
Sec.\ref{renorm} that
one can choose a suitable renormalization of the unphysical sector 
({\it heavy mass} procedure)
which avoids the appearance of two-loop vertices at this order of the 
calculation. In this case   
$\Delta \hat{\eta}_f^{(2)}$ is  given exclusively by the two-loop
wave function renormalization factor of the $Z^0$ boson 
\be
\Delta\hat{\eta}_f^{(2)}\equiv
\Delta\hat{\eta}^{(2)}= \left. {\rm Re} A^{'(2)}_{\smallz\smallz}(\mz^2)
\right|_{\msbar}~~.
\label{eta2ms}
\ee
We recall  that  the one-loop correction $\Delta\hat{\eta}_f^{(1)}$ and 
the one-loop $\msbar$ coupling counterterms do not contain $\mt^2$ terms.
Therefore $\Delta\hat{\eta}_f^{(2)}$ can be directly obtained  from
the irreducible diagrams  contributing
 to the two-loop self-energy of the $Z^0$ plus the mass counterterm graphs. 
In appendix A the sum of these two contributions to 
$A^{(2)}_{\smallz\smallz}(q^2)$ is presented. 
The explicit expression of $\Delta\hat{\eta}^{(2)}$ for arbitrary
$\mu$ can be obtained differentiating Eqs.\,(\ref{azzlh},\ref{azzhh}) 
with respect to $q^2$ and
is reported  in \equ{eta2}  (for the case $\mh \ll \mt$) and \equ{eta2b}
($\mh \gg \mw$) of  appendix B. Numerically, $\Delta\hat{\eta}^{(2)}$
is  small in all the relevant range of $\mh$ and $\mt $ values.
For $\mt=175$ GeV, $\mu=\mz$, and $\scur=0.2315$, $\Delta\hat{\eta}^{(2)}$
amounts to   $7\times 10^{-5}$ at $\mh=50$ GeV and decreases almost
monotonically for larger $\mh$. 
The two expressions for light and heavy Higgs match very
well around $\mh=96$ GeV, where they reach 3$\times 10^{-5}$.
In the range 200 GeV $\lsim\mh\lsim$ 
1 TeV, $\Delta\hat{\eta}^{(2)}$ is always less than 2$\times 10^{-5}$.

We consider now  the $Z^0$  partial widths in the
OS scheme. To derive the relevant expressions we follow the same steps and 
employ the same notation as in \efe{dgs}. The idea is to use the known
one-loop relations between the $\msbar$ and OS schemes \cite{dfs} in order 
to translate  the $\msbar$ quantities $\ecur$ and $\shat$ appearing in
\equ{gamma} into OS parameters. 
Following \efe{dgs}, we write
\bea 
\ecur^2 / \scur & = & ( G_\mu / \sqrt{2} ) 8 \mw^2 
           [ 1 - (\ecur^2 / \scur) \hat f \,], \label{gmus} \\
\scur/s^2 &  = & 1 + (c^2/s^2) (\hat{e}^2/\scur) 
\ \Delta\hat{\rho},\label{tre}
\eea
where $s^2=1-c^2 $ is an abbreviation for $\sin^2 \theta_\smallw\equiv
1-\mw^2/\mz^2$ and 
\bea 
\hat{f}& = &
\left[({\rm{Re}}
 \aww(\mw^2) - \aww(0))/\mw^2 + V_\smallw + \mw^2 B_\smallw
\right]_{\msbar},\label{sei} \\
\Delta\hat{\rho} &= &
 {\rm Re}\,\left[ \aww(\mw^2)- \ccur \azz(\mz^2)\right]_{\msbar}/\mw^2~~.
\label{quattro}
\eea
In these expressions 
$\aww(q^2)$ and $\azz(q^2)$ are the transverse $W$ and
 $Z^0$ self-energies \cite{si80,dfs,ms80} with  $\hat{e}^2/\scur$ 
factored out\footnote{We have excluded a factor $(A_{\gamma\smallz}(\mz^2))^2$
from
the $Z^0$ mass counterterm because it does not contribute at the order of the
calculation. In the numerics, however, the square of the imaginary part is not
negligible because of the additive contribution of all the light fermions,
is gauge independent, and is enhanced by a factor $\pi^2$. 
As in the case of the effective sine
\cite{gs}, we have included it  in our numerical studies.}.
 Similarly, a factor $\hat{e}^2/\scur$ has been extracted 
from the definition of $\Delta\hat{\rho}$ and $\hat{f}$.
The term $V_\smallw + \mw^2 B_\smallw$ in \equ{sei}
represents vertex and box diagram contributions to the $\mu$ decay, modulo
a factor $e^2/\scur$ \cite{si80}. 
The quantities $\hat f$ and 
$\Delta\hat{\rho}$ are actually functions of $\scur$.  
Iterating \Eqs{gmus}{tre} we can  express $\ecur^2 / \scur$ and
$\scur$ completely in terms of 
 $G_\mu$ and $s^2$. 
We note from \equ{tre} that $\scur-s^2= c^2 (\hat{e}^2/\scur)
\Delta\hat{\rho}\approx 3 \,c^2\, x_t + ...$
where
$
x_t= G_\mu \mt^2/8\pi^2 \sqrt{2} 
$
and the ellipses represent subleading  contributions. Thus, the 
replacement
$\scur= s^2\, [1 + (c^2/s^2)\,3\,x_t+ ...]$ in the one-loop 
components of $\hat{f}(\scur)$ and 
$\Delta\hat{\rho}(\scur)$ induces additional contributions of \gd. 

We recall that in \efe{dgs} two ways of writing the OS 
corrections $\Delta r$ \cite{si80} and $k$ \cite{ms80} were discussed.
The two formulations  are equivalent to \gd\, but include 
in different  ways higher orders effects. The first one,
indicated as OSI, generalizes the Consoli Hollik Jegerlehner
formula \cite{chj} for $\Delta r$ to \gd\ by writing  
\be
1-\dr= \left( 1+ \left.\frac{2\delta e}{e}\right|_{\msbar} -
\frac{e^2}{s^2} \bar{f}(s^2)\right)
\left(1+ \frac{c^2}{s^2} \frac{8 \mw^2 G_\mu}{\sqrt{2}}
\Delta\bar{\rho}(s^2) \right)\label{dieci},
\ee
where $\bar f(s^2)$ and $\Delta\bar\rho(s^2)$ are the OS counterparts
of \Eqs{sei}{quattro} and are defined in appendix B.
\equ{dieci} shows a structure for $\Delta r$ that is very close to that 
obtained in the $\msbar$ framework \cite{dfs}. 
As the $\msbar$ formulation automatically resums reducible contribution and
therefore includes part of higher order corrections, \equ{dieci}
provides an analogous resummation 
formula in the OS scheme. Moreover, similarly to $\msbar$, 
\equ{dieci} contains some residual $\mu$-dependence at $O(g^4)$.

A similar analysis can be performed for the $Z^0$ partial widths. 
Referring to \equ{gamma},  we note that the OS analogue $k_f$ 
of the form  factor $\kf$ depends on the fermion $f$, unlike its $\msbar$
counterpart.  In \efe{dgs} $k_f$ 
has  been presented only for a leptonic final state. In appendix B 
we generalize  this result   to the  case of arbitrary fermion
but the $b$ quark.   
We still have to discuss the translation of the prefactor
into the OS language.
Indeed, we can  rewrite
$\frac{\ecur^2}{\scur \ccur}|\hat{\eta}_f|$ in \equ{gamma} in terms
of OS quantities  keeping terms up to  \gd\ 
and also try to resum some higher order effect, in analogy to the 
$\msbar$ scheme. Using \Eqs{gmus}{tre},
we obtain
\be
\frac{\ecur^2}{\scur \ccur}|\hat{\eta}_f| 
\ \stackrel{OSI\ }{\longrightarrow} 
\ 4 \sqrt{2} G_\mu\mz^2
\frac{\left|
1+ \frac{  8  G_\mu \mw^2}{\sqrt{2}} \Delta\bar{\eta}_f(s^2)
\right|}
{1-\frac{  8  G_\mu \mw^2}{\sqrt{2}}
\left(\Delta\bar\rho(s^2) -\bar{f}(s^2)\right) } ,
\label{prefactor} 
\ee
 where
\be
\Delta\bar\eta_f(s^2)= \Delta\hat{\eta}_f^{(1)}(s^2) 
+\Delta\bar\eta^{(2)}_{f}(s^2)
\label{etadef}
\ee
with
\be
\Delta\bar\eta^{(2)}_{f}(s^2)= \Delta\hat{\eta}_f^{(2)}(s^2) +
\Delta\bar\eta^{(2)}_{f,add}(s^2)~~.
\label{etadef2}
\ee
In the spirit of the OSI scheme, 
\equ{prefactor} follows closely the $\msbar$ formulation, implies some
resummation, and is actually $\mu$-dependent at $O(g^4)$. 
Therefore, we adopt \equ{prefactor} whenever we employ OSI to calculate 
$s^2$.  
Explicitly, $\Delta\hat{\eta}_f^{(1)}(s^2)$ is the one-loop OS result
with  the overall coupling written in terms of $ G_\mu $,  
$\Delta\hat{\eta}_f^{(2)}(s^2)$ is given by \eqs{eta2} and (\ref{eta2b})
 in which $\scur$ has been replaced by $s^2$,
while $\Delta\bar{\eta}^{(2)}_{f,add}$ is the term induced by the shifts
$\scur\to s^2$ in the one-loop result and it is provided in appendix B.

In the second implementation of the OS scheme introduced in \efe{dgs}, called 
OSII, $\dr$ contains only the two-loop contributions proportional to 
$\mt^4$ and $\mt^2$, without any $\msbar$-like resummation, and 
it is strictly $\mu$ independent. In this case
\be
\Delta r = \Delta r^{(1)}
 + \Delta r^{(2)}  
  + \left( \frac{c^2}{s^2}\right)^2 
N_c\, x_t \left( 2\frac{e^2}{s^2} \Delta\bar{\rho}^{(1)} -
N_c\frac{\alpha}{16\pi \,s^2} \frac{\mt^2}{\mw^2}\right)
\label{quindici},
\ee
where $\dr^{(1)}$ is the original one-loop OS result of \efs{si80}{ms80},
expressed in terms of $\alpha$ and $s^2$, $\dr^{(2)}=
(e^2/s^2)\bar{f}^{(2)} - (c^2/s^2)(e^2/s^2)\Delta\bar{\rho}^{(2)}$.
The last term in \equ{quindici}
represents higher order reducible contributions induced by 
resummation of
one-loop corrections, while $\dr^{(2)}$ contains the corresponding irreducible 
components. 
In the same spirit of what has been done for
$\Delta r$ in the OSII formulation,   
we now consider a different way to write the prefactor 
$\frac{\ecur^2}{\scur \ccur}|\hat{\eta}_f|$ 
in the OS scheme.
We expand the r.h.s of \equ{prefactor} 
and  retain only two-loop effects enhanced by powers of $\mt$, 
obtaining a $\mu$-independent result:
\bea
\frac{\ecur^2}{\scur \ccur}|\hat{\eta}_f| 
& \stackrel{OSII \ \ }{\longrightarrow} \  
\frac{8 G_\mu\mz^2}{ \sqrt{2}}& \left[ 1+ 
\frac{8  G_\mu \mw^2}{\sqrt{2}}
\left( \Delta\bar{\eta}_f - \bar{f} + \Delta\bar{\rho}
\right) \right.   \nonumber \\ 
& & \left.  ~~+\frac{  8  G_\mu \mw^2}{\sqrt{2}} 
  N_c x_t
\left( 2\Delta\bar{\rho}^{(1)} +\Delta\bar{\eta}_f^{(1)} -
2\bar{f}^{(1)}
\right)- N_c^2 x_t^2\right]~~.
\label{pre2}
\eea

The two OS approaches that we have introduced 
are equivalent at \amtd but differ by 
$O(g^4)$ and higher order terms. Together with the $\msbar$ scheme,
they provide us with a broad spectrum 
of possibilities to study the scheme dependence of our results.
Summarizing our discussion of the OS scheme,
we write  \equ{gamma} as
\be
\Gamma_{f}= N_c^f\,  \frac{\mz}{96\pi} P
 \left\{ 1 - 4 I^3_f Q_f s^2 {\rm Re} k_f + 8 Q_f^2 s^4
|k_f|^2 \right\} + O(m_f^2/\mz^2),
\ee
where we evaluate the various terms  in two different ways:
(i) in the OSI approach
the prefactor $P$ is given by r.h.s of \equ{prefactor}, while the OS sine is
computed from $\alpha,\, G_\mu$ and $\mz$ using $\Delta r$ from
\equ{dieci}. The $k_f$ employed is accordingly given by 
Eq.(16) of \efe{dgs}.
(ii) in the  OSII  scheme, instead, the result  is  obtained  using \equ{pre2}
for $P$,  $\Delta r$ given by \equ{quindici}, and the factor $k_f$ given by 
Eq.(17) of \efe{dgs}.

\def \gev  {\mbox{ GeV}}
\renewcommand{\arraystretch}{1.2}
\begin{table} 
\[
\begin{array}{|c|c|c|c|}\hline
& \multicolumn{3}{|c|}{\Gamma_\ell ({\rm MeV})} \\\hline 
\mh  & {\rm OSI} & {\rm OSII}  & \msbar 
 \\  \hline\hline
65 &84.030 &84.028 &84.029    \\ \hline
100  &84.014  &84.012 &84.013   \\ \hline
300  & 83.926&83.922 &83.925    \\ \hline
600  & 83.849 & 83.845 &83.849  \\ \hline
1000 &83.791 &83.786 & 83.791 \\ \hline 
\end{array}            
\]
\caption{\sf 
Predicted values of  $\Gamma_\ell$ in OSI, OSII, and $\msbar$.
QCD corrections based on $\mut$-parametrization optimized. 
$\mt=175$GeV, $\as(\mz)=0.118$, $\Delta\alpha_{had}=0.0280$,$\mz=91.1863$.
}\label{pippot}
\end{table}

\renewcommand{\arraystretch}{1.2}
\begin{table}[t] 
\[
\begin{array}{|c|c|c|c|}\hline
& \multicolumn{3}{|c|}{\Gamma_\ell ({\rm MeV})} \\\hline 
\mh  & {\rm OSI} & {\rm OSII}  & \msbar 
 \\  \hline\hline
65 &84.051 &84.062 &84.044    \\ \hline
100  &84.035   &84.045 &84.028   \\ \hline
300  & 83.942&83.949 &83.937    \\ \hline
600  & 83.862 & 83.867 &83.858  \\ \hline
1000 & 83.801&83.804 & 83.797 \\ \hline 
\end{array}            
\]
\caption{\sf 
As in Table \ref{pippot}, but excluding $O(g^4 \mt^2/\mw^2)$ corrections. 
}\label{ta2}
\end{table}

Tables \ref{pippot} and \ref{ta2} present numerical results
for the leptonic width of the $Z^0$, $\Gamma_\ell$,
 after and before the inclusion of the \amtd
contributions. These tables  are  the 
analogue for  $\Gamma_\ell$  of the  corresponding 
tables of \cite{dgs} for $\mw$ and $\seff$. We have used 
exactly the same inputs of that paper and employed the same procedure in
deriving the numerical results. We have neglected 
 all final state mass effects.
The two-loop \cite{qcd,FKS} and leading three-loop \cite{qcd3} 
QCD corrections are calculated according to the $\mu_t $ parametrization
\cite{dgs}, a procedure of implementing the QCD
corrections in which the pole top-quark mass $\mt$ is expressed in
terms of $\hat{m}_t (\mu_t) = \mu_t$, the $\msbar$-parameter, leading
to sharply reduced QCD effects, and $\mu_t/\mt$ is evaluated by
optimization methods \cite{sirqcd}. We have verified that the 
conventional implementation of the QCD  corrections in terms of the top pole
mass leads to very close results, within 4 KeV at most.
From Tables  \ref{pippot} and \ref{ta2} we see that the scheme dependence
of the results is strongly decreased by the inclusion of the \amtd\
contributions, a pattern that was already observed for $\mw$ and $\seff$
in \efe{dgs}. In particular, we find  a maximal variation of 5 KeV 
in Table \ref{pippot}, compared to 18 KeV in Table \ref{ta2}. 
A reasonable 
estimate of the theoretical uncertainty due to uncalculated \ew\ effects 
is therefore  $O( 5)$ KeV.
Adding the QCD uncertainty,  estimated
to be  $\approx 7$ KeV in \cite{barcelona}, there is  good agreement with 
the results in Table 10 of the second paper in \cite{gattovolpe}.
We also observe good agreement between this last reference and 
  the  error estimates for $\mw$ and $\seff$ presented in 
\cite{dgs,barcelona}.
In the case of the leptonic widths,  however, the impact of these corrections
(at most 34 KeV) remains in all schemes well below the present experimental
error of 90 KeV (assuming lepton universality). 
We have also observed a reduction of the scale dependence of the $\Gamma_\ell$
prediction in the 
$\msbar$ and in the OSI scheme when the \amtd contributions are included.

\section{Discussion}
\label{discussion}
The results of the \gd\  calculations, together with all relevant 
$O(\alpha\alpha_s)$ \cite{qcd} and the leading $O(\alpha\alpha_s^2)$
QCD corrections \cite{qcd3}, 
can be summarized by very simple interpolating formulas. 
Indeed, the relative uncertainty of most of the input parameters
is quite small and their effect  on the observables can be easily linearized,
with the only exception of the Higgs mass. 
In \efe{dgps} formulae for $\seff$ and   $\mw$   as functions of
$\mh$, $\mt$, $\alpha_s$ and  the hadronic contribution 
to the running of the electromagnetic coupling 
have been presented for the three different \ew
schemes introduced in the previous sections, namely $\msbar$, OSI, and OSII.
 Here we apply  the same
procedure to the case of $\Gamma_\ell$ and slightly update the results for
 $\mw$ and $\seff$.  
The only  differences with respect to \efe{dgps} are  that here we employ
(i) the new value  $G_\mu=1.16637\times 10^{-5}$ GeV$^{-2}$ 
obtained after incorporation 
of the complete two-loop QED corrections to the muon decay
\cite{stuart} and (ii) the new central value $\mz=91.1867$ GeV.
These two changes in the inputs
  amount  to  a very small shift of the overall 
normalization, common to the three schemes.
The analytic formulae are of the form
\be
        \seff  =  (\seff)_0 +c_1\, A_1 + c_2\, A_2 - c_3\, A_3 + c_4\, A_4,  
                                                        \label{eq:s2}
\ee
\be
        \mw    =  \mw^0   - d_1\, A_1 - d_5\, A_1^2 - d_2\, A_2 + d_3\, A_3 
                - d_4\, A_4,                              \label{eq:Mw}
\ee
\be
  \Gamma_\ell  =  \Gamma_\ell^0 - g_1\, A_1 -g_5\, A_1^2- g_2\, A_2 +
 g_3\, A_3 - g_4\, A_4,  
                                                        \label{eq:gl}    
\ee
where 
\bea
     &   A_1  = \ln \frac{\mh}{ 100\gev}, \ \ \ \ \ \ \ \ \ \ \ \ \ \        
     &    A_2  =  \frac{ (\Delta \alpha)_h }{ 0.0280} -1, \nn\\   
     &    A_3  =  \left(\frac{\mt}{175 \gev}\right)^2 -1, \ \ \ \ \ \ \ 
     &    A_4  =  \frac{\alpha_s(\mz)}{0.118} -1,\nn
\eea
$(\Delta \alpha)_h$ is the five-flavor hadronic contribution to the QED
vacuum-polarization function at $q^2=\mz^2$, and $(\seff)_0$, $\mw^0$,
and $\Gamma_\ell^0$ are (to excellent approximation) 
the theoretical results at the reference point $\mh=100$ GeV,
$(\Delta \alpha)_h=0.0280$, $\mt = 175$ GeV, and $\alpha_s(\mz)=0.118$. 
The values of $(\seff)_0$, $\mw^0$, $\Gamma_\ell^0$, $c_i$ 
$(i=$1-4) and 
$d_j,g_j$ $(j=$1-5) for the three electroweak schemes of Ref.\cite{dgs}
 are given in Tables \ref{t:seff}-\ref{t:gl}. 
We present  the coefficients in the case of the 
$\mu_t$-parametrization  mentioned in the previous section, which was shown in
\efe{dgs} to give results very close to the more conventional 
parametrization in terms of the top pole mass. 
These formulae are very  accurate for $75 \lsim\mh\lsim 350$ GeV with the
other parameters in the ranges
$170\lsim\mt\lsim 181$ GeV, $0.0273\lsim\Delta\alpha_h\lsim
0.0287$, $0.113\lsim\as(\mz) \lsim 0.123$. In this case they reproduce 
the exact results of the calculation with maximal errors of 
$\delta s^2_{eff}\sim 1\times 10^{-5}$,
$\delta\mw\lsim 1$ MeV and  $\delta\Gamma_\ell\lsim 3$ KeV, which are all very 
much below the experimental accuracy. Outside the above range, the
deviations increase but remain very small for larger Higgs mass, 
reaching about $3\times 10^{-5}$, 3 MeV, and 4 KeV
at $\mh=600$ GeV for $\seff$, $\mw$, $\Gamma_\ell$, respectively.
\renewcommand{\arraystretch}{1.15}
\begin{table}[t]
\begin{center}
\begin{tabular}{|c||c|c|c|c|c|} 
\hline 
 Scheme & $(\seff)_0$  & $10^4 c_1$  & $10^3 c_2$  & $10^3 c_3$ & 
        $10^4 c_4$  \\ \hline \hline 
$\msbar$          &0.231513 &  5.23 & 9.86 & 2.78 & 4.5    \\ \hline 
OSI            &0.231527 &  5.19 & 9.86 & 2.77 & 4.5    \\ \hline
OSII           &0.231543 &  5.26 & 9.86 & 2.68 & 4.4    \\ \hline
\end{tabular} 
\caption{Values of $(\seff)_0$ and $c_i$  in Eq.~(\ref{eq:s2})
        for three electroweak schemes that incorporate $O(g^4\mt^2/\mw^2)$
        corrections in the $\mu_t$-parametrization of QCD corrections 
        [2].
}
\label{t:seff}
\end{center} 
\end{table} 
\begin{table}[t]
\begin{center}
\begin{tabular}{|c||c|c|c|c|c|c|} 
\hline 
 Scheme & $\mw^0$  & $10^2 d_1$  & $10 \,d_2$  & $10\, d_3$ & 
         $10^2 d_4$  & $10^3 d_5$ \\ \hline \hline 
$\msbar$          & 80.3829 &  5.79 & 5.17 & 5.43 & 8.5 & 8.0    \\ \hline 
OSI            & 80.3809 &  5.73 & 5.18 & 5.41 & 8.5 & 8.0      \\ \hline 
OSII           & 80.3807 &  5.81 & 5.18 & 5.37 & 8.5 & 7.8      \\ \hline 
\end{tabular} 
\caption{Values of $\mw^0$ and $d_i$ in Eq.~(\ref{eq:Mw}),
        in GeV,  as in Table \ref{t:seff}.}
\label{t:mw}
\end{center} 
\end{table} 
\renewcommand{\arraystretch}{1.15}
\begin{table}[t]
\begin{center}
\begin{tabular}{|c||c|c|c|c|c|c|} 
\hline 
 Scheme & $\Gamma_\ell^0$  & $10^2 g_1$  & $10 g_2$  & $10 g_3$ & 
        $10 g_4$ & $10^2 g_5$ \\ \hline \hline 
$\msbar$          &84.0112 &  5.37 & 4.76 & 8.01 & 1.13 &2.14   \\ \hline 
OSI            &84.0122 &  5.39 & 4.76 & 8.02 & 1.12 & 2.16   \\ \hline
OSII           &84.0099 &  5.49 & 4.75 & 7.99 & 1.12 & 2.16   \\ \hline
\end{tabular} 
\caption{Values of $\Gamma_\ell^0$ and $g_i$  in Eq.~(\ref{eq:gl}),
        in MeV, as in Table \ref{t:seff}.
} \label{t:gl}
\end{center}
\end{table} 

By comparing the normalized coefficients of $\ln \mh$ in \eqs{eq:s2} and 
(\ref{eq:Mw}), we see that $\seff$ is $\sim 3$
times more sensitive to $\ln \mh$ than $\mw$, $\sim 6.6$ times more sensitive
to $\Delta\alpha_h$, almost 2 times more sensitive to $\mt$ and $\as$.
Concerning $\Gamma_\ell$, table \ref{t:gl} shows a sensitivity to  $\ln \mh$
comparable to that of $\mw$. Its present experimental determination, however,
is less precise and it is not expected to improve significantly in the
foreseeable 
future. Despite the recent progresses in the measurement of $\mw$,
 it is  clear that most of the present
sensitivity to $\mh$ still comes from the effective sine. However, with an 
improvement in the measurements of both $\mw$ and $\mt$
the situation may change in the  future \cite{dgps,peppe,barcelona}.

We now use Eqs.(\ref{eq:s2}--\ref{eq:gl}) together with the latest data
to determine $\mh$ by a $\chi^2$ analysis. 
The experimental inputs  we use are \cite{LEPEWWG}:
$\seff=0.23157\pm 0.00018$, $\mw=80.394\pm 0.042$ GeV, 
$\Gamma_\ell=83.90\pm 0.10$ MeV, $\mt=174.3\pm 5.1 $GeV,
$\as(\mz)=0.119\pm0.003$. Concerning 
 $(\Delta\alpha)_h$ we adopt  the value 
$0.02804\pm0.00065$ \cite{jeg} but will comment later on the impact of some
recent  estimates \cite{jeger}. 
The dependence on the scheme  of our result is very mild. 
The $\msbar$ scheme leads to the highest
$\mh$ value. The  $\ln (\mh/100)$ is to  good
approximation normally distributed and we obtain in $\msbar$ scheme
$\ln \mh/100=-0.04\pm0.64$, corresponding to a central value $\mh=96$
GeV and to a 95\% probability upper limit of  about 270 GeV.
Repeating the same analysis in the other two schemes, we obtain 
95\% upper limits that differ from the one in $\msbar$ by at most $-15$ GeV. 
One should also consider 
the uncertainty  due to uncalculated QCD corrections. The dominant QCD effect
is linked to the leading $O(G_\mu \mt^2)$ contribution to  $\Delta\rho$,
and in particular to the top quark mass definition. A simple way
to take the related uncertainty into account is through a systematic
shift of $\mt$. Indeed, the conservative 
values for the QCD uncertainty given in \cite{dgps,barcelona}
 correspond to a variation of 
$\delta \mt=\pm 0.9$ GeV. An increase of 0.9 GeV in the central value of
the top mass moves the $\mh$ 95\% upper bound 
by about $+15$ GeV. Thus $\mh \lsim 285$ GeV could be taken as a 95\%
probability upper limit obtained using only indirect information and  
including the theory uncertainty.
Finally, we note that a corresponding analysis
carried out without implementing the $O(g^4 \mt^2/\mw^2)$ corrections 
gives a central value and  upper bound for $\mh$  significantly
larger, by about 40 and 90 GeV, respectively.

Following \efe{higgsbound}, we can include in the $\mh$ analysis the 
information coming from the direct search experiments presently in operation 
at LEP. The combination of the search data, for center of mass
energy up to $\sqrt s = 183 $ GeV, with the above result from precision 
measurements moves the central value for $\mh$ from 96 to 145 GeV,
while the 95\% probability upper limit including the theory uncertainty
reaches 345 GeV. 
The uncertainty on the upper bound due to the simplified likelihood procedure 
adopted has been estimated in \cite{higgsbound} to be $O(5\,$GeV).

Because of the strong correlation between
$\mt$ and $\mh$, apparent in Eqs.(\ref{eq:s2}-\ref{eq:gl}), 
in a global fit also observables 
insensitive to the Higgs boson can have an indirect effect in the 
$\mh$ determination. In particular, the prediction for $R_b$ is almost 
independent of $\mh$, but the present experimental value
$R_b=0.21680\pm0.00073$ 
points towards  a much lighter top ($\mt\approx140\pm 24$ GeV)
than most other data. A smaller $\mt$ central value favors a lighter 
$\mh$  and consequently leads to a reduced 95\% upper bound. The inclusion of
$R_b$ in the fit, for example,
would lower the $\mh$ central value and upper limit by approximately  10 and 
30 GeV, respectively. 
Our analysis, in any case,  is based only on 
$\mw$, $\seff$, and $\Gamma_\ell$,  the three best measured observables
which are also   known to \gd.

Several new estimates of  the hadronic contribution to the running of 
$\alpha$ have appeared over  the last couple of years \cite{jeger}. These
new investigations differ from the previous most phenomelogical
analyses \cite{jeg} mainly by the use of perturbative QCD (pQCD) down to 
energies of the order of a few GeV and by the treatment of 
old experimental data in regions where pQCD is not applicable. 
The combination of these two factors gives a result  for $(\Delta\alpha)_h$
with a drastically reduced uncertainty but, at the same time, with a lower
central value. Since these new estimates remove a significant part of the 
theoretical error in the prediction of $\seff$, they further
enhance  its role 
in the indirect $\mh$ determination. As can be seen from the
sign of the coefficient $c_2$ in \equ{eq:s2}, a smaller central value
for $(\Delta\alpha)_h$ prefers a heavier Higgs. 
There is therefore a compensation between the
reduced error and the lower central value so that 
only  the very high confidence level values are significantly reduced.
As an illustration, we perform the $\mh$ analysis using the $\Delta\alpha_h$
 determination with the
smallest error, due to Davier and Hocker, i.e. $\Delta\alpha_h=0.02770\pm
0.00016$. The $\msbar$ fit, without including theoretical errors and the
information from direct searches, has a central value at $\mh=122$ GeV and an
upper bound only  10 GeV smaller than  the one obtained with 
the  $\Delta\alpha_h$ determination of \cite{jeg}.

\section{Summary}
The \amtd corrections have  been calculated for many accurately measured 
precision observables and are now implemented in several global fit programs
\cite{topaz0,zfitter,erler}. 
The only important quantities whose predictions are not 
known at the same order are those connected with $b$-final states because
their \gd\ contributions involve specific two-loop vertex 
corrections.

In this paper we have presented the results of the calculation of the 
partial widths $\Gamma_f$ ($f\neq b$) of the $Z^0$ boson to \amtd in three
different \ew renormalization schemes.
In the case of $\Gamma_\ell$, 
the size of these contributions is  modest (up to 34 KeV) 
when compared to the present experimental error (90 KeV),
but,  like for  $\mw$ and the effective sine, their inclusion 
sharply reduces the scheme dependence of the theoretical predictions.
The situation is very similar for the partial widths into quarks.
Concerning $\Gamma_u$ and $\Gamma_d$, in the case of
 a light Higgs of $100$ GeV  the \gd\ corrections induce the  variations  
$\Delta \Gamma_u = -63$ KeV and  $\Delta \Gamma_d = -85$ KeV in the  
$\msbar$ scheme, $\Delta \Gamma_u = -62$ KeV and $\Delta \Gamma_d = -73$ KeV 
in OSI, while the shifts in OSII are approximately twice larger,
$\Delta \Gamma_u \simeq\Delta \Gamma_d = -175$ KeV. It is interesting to 
compare these shifts with the ones due to non-factorizable 
$O(\alpha \alpha_s)$ corrections to the hadronic width: 
$\Delta \Gamma_u = -124$ KeV, $\Delta \Gamma_d = -173$ KeV \cite{oronzo}.
Adding the contributions of the $u,d,s,c$ quarks, the overall shift 
to the total hadronic width $\Gamma_h$ may reach 0.7 MeV, to be 
compared to the present experimental error of 2.3 MeV.

We have also described in detail several aspects of the \amtd\ calculations, 
which had not been reported in previous publications. 
Furthermore, in appendix A we have presented explicit expressions for the
two-loop \gd\ contribution to the 
self-energies of the $W$ and $Z^0$ bosons.   

The analytic results of the calculations for 
$\mw$, $\seff$, and $\Gamma_\ell$, together with the
relevant  QCD corrections, have been incorporated in a numerical code,
from which  we have derived  simple 
interpolating formulae following \efe{dgps}. They 
reproduce very accurately the  results of the code
for a large range of values of the input parameters.
At present, one of the most interesting application of  high precision
investigations is the attempt to constrain the Higgs boson mass from the 
data. As has been repeatedly observed \cite{dgs,dgps,peppe,barcelona}, 
our \amtd\ calculations do have a significant impact on the Higgs mass bound,
 because of an enhancement of the {\it screening} operated by the  
higher order contributions on the leading one-loop correction.
Using the interpolating formulae, it is very simple to obtain an indirect 
determination of  $\mh$ based on the latest preliminary data.
The result of our analysis using only the indirect information
from precision measurements gives $\mh < 285$ at 95\% C.L.,
where  the theoretical uncertainty is also taken into account.
The inclusion of the direct search information up to energies $\sqrt s =
183$ GeV according to the procedure outlined in \cite{higgsbound}
increases the  95\% upper limit by $\sim 60$ GeV. 
As a final remark,
we notice that the Higgs determination is very sensitive to the value of 
the effective sine employed in the analysis and that the above results have
been obtained using the SLD+LEP average. Although recently the 
discrepancy between  the LEP and SLD $\seff$ determinations 
has  somewhat receded, it should be kept in mind that
their agreement is  still far from satisfactory. 

\subsection*{Acknowledgements}
We are grateful to S.~Fanchiotti, F.~Feruglio, 
M.~Passera, A.~Sirlin, and A.~Vicini for collaboration 
on different projects variously  connected to the subject of this paper.
We also thank P.A. Grassi, G. Passarino, and M. Steinhauser
for interesting communications and discussions.
This work has been supported in part 
by the Bundesministerium f{\"u}r Bildung  und
Forschung under contract 06 TM 874 and by the DFG project Li 519/2-2.

\renewcommand{\thesection}{\Alph{section}}
\begin{appendletterA}
\section*{Appendix A}
\label{appA}
\appendix
In this  appendix we present  some of the 
intermediate results. The analytic expressions for the irreducible 
two-point functions of the $W$ and the $Z^0$  bosons through \gd\ 
are extremely long and cannot be reduced to a compact form.
However, after on-shell renormalization of the masses inside the loops, 
i.e. after adding the mass counterterm diagrams, they simplify considerably.

The expressions for the two-loop self-energies that follow are
calculated in the 't Hooft-Feynman gauge and include all the irreducible 
unrenormalized diagrams that contribute to the order of the calculation
plus the corresponding contributions from the on-shell mass counterterms
of the particles in the loops.
  Neither gauge coupling nor wave-function renormalization are carried out.
The renormalization of the unphysical $Z^0$ and $W$ sectors
 is performed according to the {\it heavy mass}
 procedure described in Sec.\,\ref{renorm},  so that  all two-loop
vertex and box contributions proportional to powers of $\mt$ are canceled, 
and only two-loop self-energies have to be taken into account.

As the couplings of the SM sometimes contain masses, which we 
renormalize      in the same way as the ones that appear in the propagators, 
a possible ambiguity arises in the $\msbar$ scheme, because 
it is always possible to rewrite $\mw$ as $\mz \chat$ in the 
couplings of a one-loop diagram;
 the counterterm associated to the two forms 
 differ in the finite part.  In order to avoid any possible 
source of confusion, it is important to stress here that all our two-loop 
expressions are therefore consistent
only with the choices adopted in the one-loop $\msbar$ expressions reported in 
\cite{dfs,ds91}. Different choices would 
induce different \gd\  terms.

We first introduce some functions which appear in 
the following  expressions of $\aww(q^2)$ and $\azz(q^2)$ and in the formulae
of the next appendix.
We define:
\bea
g(x) =  \left\{
          \begin{array}{lr}
           \sqrt{4-x}\,\left
              (\pi - 2 \arcsin{\sqrt{x/4}}
                       \right) &                        0 < x \leq 4 \\
           {}\\
           2 \sqrt{x/4-1}\,\ln\left(
                \frac{1-\sqrt{1-4/x}}{1+\sqrt{1-4/x}}
                              \right) &                 x > 4 \,~ ,
          \end{array}
        \right. \nn 
\eea
\bea
\Lambda( x) =  \left\{
          \begin{array}{lr}
          -\frac{1}{2 \sqrt{x}}\,g(x) + {\pi \over 2} \,\sqrt{4/x -1}
                      &                         0 < x \leq 4 \\
           {}\\
            -\frac{1}{2 \sqrt{x}}\, g(x)&               x > 4 \,~ ,
          \end{array}
        \right.  \nn 
\eea
where we have indicated the dilogarithmic function as 
${\rm{Li_2}} (x) = - \int_0^x dt \, {\ln (1-t) \over t} $, 
and we introduce 
\bea
\phi(z) =\left\{
       \begin{array}{ll}
       4 \sqrt{{z \over 1-z}} ~Cl_2 ( 2 \arcsin \sqrt z ) &  0 < z
\leq 1\\
       { 1 \over \lambda} \left[ - 4 {\rm Li_2} ({1-\lambda \over 2}) +
       2 \ln^2 ({1-\lambda \over 2}) - \ln^2 (4z) +\pi^2/3 \right] 
        &z >1 \,,
       \end{array}\right.\nonumber
\eea 
where $Cl_2(x)= {\rm Im} \,{\rm Li_2} (e^{ix})$ is the Clausen function
and $\lambda = \sqrt{1 - {1 \over z}}$.

The function  $B_0[q^2,m_1^2,m_2^2]$ is defined through the one-loop 
integral ($\eps= (4-n)/2$)
\be
-i \,\mt^{2\epsilon} e^{\gamma\epsilon}\int \frac{d^n k}{\pi^{n/2}}
\frac1{[k^2-m_1^2][(k-q)^2-m_2^2]}=
\frac1{\epsilon} + B_0[q^2,m_1^2,m_2^2] + O(\epsilon),
\label{B0}
\ee
whose analytic form is well known (see for example \efe{ds92}).
 It is interesting 
to note that the $O(\eps)$ part of one-loop integrals like the one in 
\eqs{B0} cancel exactly in the two-loop expressions of the self-energies
given below.

We also introduce the following short-hand notation: 
\[
\begin{array}{|c|}\hline 
w=\mw^2;\ \ \ \ \
z=\mz^2;\ \ \ \ \ 
h=\mh^2;\ \ \ \ \
t=\mt^2;\nn\\
 w_t= \frac{w}{t}; \ \ \ \ \ \ \ 
z_t= \frac{z}{t}; \ \ \ \ \ \  \ 
h_t= \frac{h}{t}. \\
\hline
\end{array}\]
As no coupling renormalization has been carried out in the following
self-energies, we express the results in
terms of unspecified sine and cosine $s,c$ of the Weinberg angle which should
{\it not} be understood as {\it on-shell} quantities.

\subsection*{$\bullet\ \azz(q^2)$}
In order to fix completely the one-loop contribution we report here
the coefficient of the $1/\epsilon$ pole term  
 that is not explicitly written in \efe{dfs}
\be
\label{azz1}
A_{\smallz \smallz}^{(1)\,pole} =
\frac{g^2}{(16\pi^2)} \left[ q^2 \left( \frac{41}{3} -
\frac{41}{6 \, c^2} - \frac{23\,c^2}3 \right) + \frac{3}{2 \,c^2} \mt^2
+ \left( 4 - \frac2{c^2} \right) \mw^2 - \frac{\mz^2}{c^2} \right]
\ee
where $\mt,\, \mw$ and $\mz$ are understood as OS masses.
Concerning the HTE of the two-loop self-energy,
as explained in Sec.\,\ref{hte},
we have to distinguish between the case in which the Higgs is light relative to
the top quark, and the one in which it is comparable to it.
In the first case, the result, expressed in 
 units $N_c g^4/c^2 \ \mt^2/(16\pi^2)^2 \ (\mu/\mt)^{4\epsilon}$ 
$(N_c =3)$, reads
\bea
\label{azzlh}
A^{(2)}_{\smallz\smallz}(q^2)&=&
\left( {{13}\over {24}} + 
{{1}\over {48\,{c^2}}} + {3\over {16\,{ w_t}}}\right)\frac1{\epsilon^2}
\nn\\&+&
\left[ \sqrt{{\it h_t}}\,\pi\frac{ 3\,h_t - 8  }{16\,w_t} -
{1\over {8\,{c^2}}} \ln z_t-
\frac1{4}  \ln w_t
+ \left({{17}\over {72}} - 
   {{17{c^2}}\over {36}} + {{13{c^4}}\over {36}}\right)\frac{q^2}{w} - {c^2} 
\right.
\nn\\&+&\left.
  {{197}\over {144}} + {{29}\over {288{c^2}}}+ 
   {{3h}\over {8\,w}} +    {{23}\over {32\,w_t}}
-{{3\,h}\over {8\,w}} \ln h_t 
\right]\frac1{\epsilon}
+{{h\,\left( {c^2}\,\left( 2\,h + 5\,{q^2} \right)  - 2\,w \right) }\over 
    {36\,{c^2}\,{q^2}\,w}}\ln h_t 
\nn\\&+&
\left({1\over 12} + {1\over {48\,{c^2}}} - {1\over {48\,w_t}}\right)\pi^2
   -{{ 81h + 136{q^2} - 320{c^2}{q^2} + 256{c^4}{q^2} - 
        216t   }\over {432\,w}}
  \sqrt{{\it h_t}}\,\pi 
\nn\\&+&
{{{c^2}( h - {q^2} )  - w}\over {18\,{c^4}\,{q^2}}}\ln c^2\
+  \left(  {1\over {18\,{c^2}}} -{1\over {24}}+ {{{c^2}}\over 3} - 
   {h\over {18{c^2}{q^2}}} + {w\over {18\,{c^4}\,{q^2}}}\right)\ln w_t
\nn\\&+&
\left(  {5\over 6} - {{2\,{c^2}}\over 3} + {{{q^2}}\over {24w}} - 
   {{{c^2}\,{q^2}}\over {12w}}
\right) B_0[q^2,w,w] 
-\frac{q^2}{18w} (1+2c^2)\ln \left(\frac{-q^2}{\mt^2}\right)
-\frac1{2} \delta_t  
  \nn\\&+&
\left(  {{ 5{q^2}-h}\over {9\,{c^2}}} + 
   {{(h-q^2)^2}\over {18\,w}} + 
   {w\over {18{c^4}}}
\right)\frac{B_0[q^2,h,z] }{q^2}
+\left(
  {{61}\over {108}} - {{191{c^2}}\over {216}} + {{61{c^4}}\over {72}}
\right)\frac{q^2}{w}  
  \nn\\&-&{{(h - z ) ^2}\over {18\,w\,q^2}}
-  {{17}\over {96}} + {{55}\over {192{c^2}}} - {{{c^2}}\over 2} - 
   {{7h}\over {48w}} - {{45 }\over {64w_t}}.
\eea
Here and henceforth 
$\delta_t$ is the $O(\epsilon)$ part of the top mass counterterm 
calculated at $\mu=\mt$
which cancels out  in the physical amplitudes we have considered. 
In the case of a heavy Higgs boson, the HTE takes the form,
 in the same units:
\bea
\label{azzhh}
A_{\smallz\smallz}^{(2)}(q^2)&=&
\left( {{13}\over {24}} + 
{{1}\over {48\,{c^2}}} + {3\over {16\,{ w_t}}}\right)\frac1{\epsilon^2}
\nn\\&+&
\left[\frac{ h_t - 4  }{16\,w_t} g(h_t) \sqrt{h_t} -
{\ln z_t\over {8\,{c^2}}}  -
\frac{\ln w_t}{4}
+ \left({{17}\over {72}} - 
   {{17{c^2}}\over {36}} + {{13{c^4}}\over {36}}\right)\frac{q^2}{w}
-c^2 +  {{197}\over {144}}
\right.
\nn\\&+&\left.
 {{29}\over {288{c^2}}} - 
   {{h}\over {8\,w}} +    {{23}\over {32\,w_t}}
+{{h_t ( h_t - 6  ) }\over {16\,w_t}} \ln h_t 
\right]\frac1{\epsilon}
-\frac{q^2}{18w} (1+2c^2)\ln \left(\frac{-q^2}{\mt^2}\right)
\nn\\&+&
\left({1\over 12} + {1\over {48\,{c^2}}} - {1\over {48\,w_t}}\right)\pi^2
+\left(  {5\over 6} - {{2\,{c^2}}\over 3} + {{{q^2}}\over {24w}} - 
   {{{c^2}\,{q^2}}\over {12w}}
\right) B_0[q^2,w,w]-\frac1{2} \delta_t 
  \nn\\&+&
\left[  {{3( h_t - 4 ) }\over {8{c^2}\,h_t}} + 
   {{h( 7-h_t ) }\over {16\,w}} +\frac{q^2}{w}\left(
{{(4c^2 -5)c^2
      ( 6 + 27{h_t} - 10{{{ h_t}}^2} + {{{\it h_t}}^3} ) }
\over {54\left( {\it h_t}-4 \right) }}\right.\right.
\nn\\&-&\left.\left.
  {{1152 + 606{\it h_t} + 1467{{{\it h_t}}^2} - 1097{{{\it h_t}}^3} + 
      238{{{\it h_t}}^4} - 17{{{\it h_t}}^5}}\over 
    {432{{\left( {\it h_t}-4 \right) }^2}\,{\it h_t}}}
\right)\right]\ln h_t
  \nn\\&+&
\left(\frac{h_t}{4}-1\right)\left[
\frac{q^2}{w}\left( 
  {{17}\over {108}} - {{10\,{c^2}}\over {27}} + {{8\,{c^4}}\over {27}}
\right)
-{{ 1  }\over {4\,w_t}}
\right] \sqrt{h_t} \,g(h_t)
-  {{17}\over {96}} + {7\over {192\,{c^2}}} 
  \nn\\&-& {{{c^2}}\over 2} + 
   {h\over {16w}} - {{77}\over {64w_t}}
+\left( {{{c^2}}\over 3}-{1\over {24}} \right)\ln w_t
+\left[ {{3\left( 4 - {\it h_t} \right) \left({\it h_t} -2\right) }\over 
     {16{c^2}\,{{{\it h_t}}^2}}} + {{3}\over {16\,w_t}}
\right.
\nn\\&+&\left.
\frac{q^2}{w}\left(
  {{{c^2}\left( 5 - 4\,{c^2} \right) \left( {\it h_t}-1 \right) }\over 
    {9\,\left( h_t-4 \right) \,{\it h_t}}}
-{{384 + 10{\it h_t} - 238\,{ h_t^2} + 63\,{ h_t^3} - 
      3\,{ h_t^4}}\over 
    {144\,{{\left(  h_t-4 \right) }^2}\,{{{\it h_t}}^2}}}
\right)
\right]\phi(\frac{h_t}{4})
 \nn\\&+&
\left(  {{3}\over {{c^2}\,h_t}} -{{3}\over {4\,{c^2}}}  - {h\over {8\,w}} - 
   {{{q^2}}\over {12\,w}} + {1\over {2\,w_t}} + {{{q^2}}\over {3\,h_t\,w}}
\right)\Lambda(h_t)
 \nn\\&+&
\frac{q^2}{w}\left[ {{\left( -17 + 40\,c^2 - 32\,c^4 \right) 
{\it h_t}}\over {216}}+  
{5\over {144\left({\it h_t} -4\right)}}+
{{791}\over {864}} - {{251\,{c^2}}\over {216}} + {{77\,{c^4}}\over {72}} 
\right].
\eea

\subsection*{$\bullet$ $\aww(q^2)$}
The coefficient of one-loop pole is
\be
\label{aww1}
A_{\smallw \smallw}^{(1)\,pole} =
\frac{g^2}{(16\pi^2)} \left[ -\frac{5}{6} q^2 + \frac{3}{2} \mt^2
-\frac{s^2}{c^2} 
 \mw^2 + \mz^2 \,c^2 \right]~~.
\ee
In the case of a light  Higgs, the HTE for the two-loop
$W$ self-energy takes the form,
in units $N_c g^4 \, \mt^2/(16\pi^2)^2\ (\mu/\mt)^{4\epsilon}$, 
\bea
\label{awwlh}
A_{\smallw\smallw}^{(2)}(q^2)&=&
\left( {{13}\over {24}} + 
{{1}\over {48\,{c^2}}} + {3\over {16\,{ w_t}}}\right)\frac1{\epsilon^2}
\nn\\&+&
\left[ \sqrt{{\it h_t}}\,\pi\frac{ 3\,h_t - 8 }{16\,w_t} -
{1\over {8\,{c^2}}} \ln z_t  - \frac1{4}  \ln w_t
+\frac{q^2}{8w}+ \frac{233}{144} + \frac{101}{288 c^2}
\right.
\nn\\&+&\left.
     {{3h}\over {8\,w}} +    {{23}\over {32\,w_t}}
-{{3\,h}\over {8\,w}} \ln h_t 
\right]\frac1{\epsilon}
+  {{h\,\left( 5\,h - {q^2} - 5\,w \right) }\over {144\,{q^2}\,w}} \ln h_t
 -\frac1{2} \delta_t 
\nn\\&+&
\left( {{23}\over {108}} - {{55}\over {216\,{c^2}}} + {h\over {8\,w}} - 
   {{{q^2}}\over {36\,w}} - {7\over {48\,w_t}}
\right)\pi^2
-{{{c^2}\,{q^2}   + s^2 w }\over {48\,{c^4}\,{q^2}}}\ln c^2
  \nn\\&+&
\left( 
 {{3w -5\,{c^4}\,h + 3{c^2}\,{q^2} - 46{c^4}{q^2}  - 6\,{c^2}\,w + 
      8\,{c^4}\,w}\over {144\,{c^4}\,{q^2}}}\right)\ln w_t 
  \nn\\&+&
\left( {{11 - 13\,{c^2}}\over {24\,{c^2}}} + {{{q^2}}\over {48\,w}} + 
   {{{s^4}\,w}\over {48\,{c^4}\,{q^2}}}
\right) B_0[q^2,w,z] +    {{31{q^2}}\over {48w}}
  \nn\\&+&
\left(5\,{q^2}-h + {{(h-q^2)^2}\over {2\,w}} + {w\over 2}
\right)\frac{5\,B_0[q^2,h,w] }{72q^2}
+{{24t -9\,h - 32\,{q^2} }\over {96\,w}}  \sqrt{{\it h_t}}\,\pi 
  \nn\\&+&
\left(   {{10h}\over w}
  -8 - {3\over {{c^4}}} + {6\over {{c^2}}} - {{5{h^2}}\over {{w^2}}} 
\right)\frac{w}{144q^2}
  -{{155}\over {288}} + {{1633}\over {576{c^2}}} 
- {{29h}\over {24w}}  + {{31}\over {64w_t}},
\eea
while in the case of a heavy Higgs, we find in the same units:
\bea
\label{awwhh}
A_{\smallw\smallw}^{(2)}(q^2)&=&
\left( {{13}\over {24}} + 
{{1}\over {48\,{c^2}}} + {3\over {16\,{ w_t}}}\right)\frac1{\epsilon^2}
\nn\\&+&
\left[ \sqrt{{\it h_t}} \,g(h_t){{h_t - 4}\over {16\,w_t}}-
{1\over {8\,{c^2}}} \ln z_t - \frac1{4}  \ln w_t
+\frac{q^2}{8w}+ \frac{233}{144} + \frac{101}{288 c^2}
\right.
\nn\\&-&\left.
     {{h}\over {8\,w}} +    {{23}\over {32\,w_t}}
+{{h_t\left( h_t - 6 \right) }\over {16\,w_t}}\ln h_t 
\right]\frac1{\epsilon}
+  {{\left(4 -h_t\right) \left( 3\,h + 2\,{q^2} + 18\,w \right) }\over 
    {24\,h\,w_t}}\Lambda(h_t)
\nn\\&+&
\left(
  {{23}\over {108}} - {{55}\over {216\,{c^2}}} + {1\over {16\,h_t}} - 
   {{{1}}\over {24\,{h_t^2}}} + {1\over {96\,w_t}} - 
   {{7\,{q^2}}\over {144\,h_t\,w}} - {{{1}}\over {16\,h_t\,w_t}} - 
   {{19\,{q^2}}\over {432\,{h_t^2}\,w}}
\right)\pi^2
\nn\\&+&
\left( {{11 - 13{c^2}}\over {24\,{c^2}}} + {{{q^2}}\over {48\,w}} + 
   {{{s^4}\,w}\over {48{c^4}\,{q^2}}}
\right) B_0[q^2,w,z]
-{{{c^2}\,{q^2}   + s^2 w }\over {48\,{c^4}\,{q^2}}}\ln c^2
\nn\\&+&
\left[  {5\over {16}} - {7\over {4\,{\it h_t}}} - {3\over {8\,{\it w_t}}} + 
   {{{\it h_t}}\over {16\,{\it w_t}}} - {{{{{\it h_t}}^2}}\over 
{32\,{\it w_t}}} +\frac{q^2}{w}\left(
  {1\over {36}} - {{31}\over {72\,{\it h_t}}} - {{{\it h_t}}\over 8} + 
   {{{{{\it h_t}}^2}}\over {24}}\right) \right]\ln h_t
\nn\\&+&
  \sqrt{{\it h_t}} \,g(h_t)\left( h_t - 4 \right) 
  \frac{4\,{q^2} - 3\,t 
}{96 \,w}
+\left(
 {{1 - 17\,{c^2}}\over {48\,{c^2}}} + {{{s^4}\,w}\over {48\,{c^4}\,{q^2}}}
\right)\ln w_t
- {{{s^4}\,w}\over {48\,{c^4}\,{q^2}}}
\nn\\&+&  \frac{ h_t -4}{288{{{\it h_t}}^2}\,w}
      \left( 12{q^2} - 34\,{\it h_t}\,{q^2} - 26\,{{{\it h_t}}^2}\,{q^2} + 
        18{{{\it h_t}}^3}\,{q^2} + 54\,{{{\it h_t}}^2}\,t - 
        27{{{\it h_t}}^3}\,t + 108\,w \right.
 \nn\\&-&
\left. 18\,{\it h_t}\,w + 
        18{{{\it h_t}}^2}\,w \right) 
\phi (\frac{h_t}{4})
+  \left(38{q^2} + 42\,{\it h_t}\,{q^2} - 24\,{{{\it h_t}}^2}\,{q^2} - 
      62{{{\it h_t}}^3}\,{q^2} + 18{{{\it h_t}}^4}\,{q^2} 
\right.
 \nn\\&+&
\left.       54{\it h_t}\,t - 135\,{{{\it h_t}}^2}\,t + 108\,{{{\it h_t}}^3}\,t - 
      27 h_t^4 t + 36w - 54{\it h_t}\,w + 18 h_t^3\,w
\right)
     {{\rm Li}_2(1-h_t) \over {144\,{{{\it h_t}}^2}\,w}}
 \nn\\&-&
  {{23}\over {72}} + {{1633}\over {576\,{c^2}}} + {1\over {4h_t}} - 
   {{3\,h}\over {16\,w}} + {{23}\over {64\,w_t}}
+\frac{q^2}{w}\left(  {5\over {36}} + {{{\it h_t}}\over {24}} + 
{{19}\over {72h_t}}
\right)-\frac1{2} \delta_t .
\eea
The other  self-energies used in the various \gd\ calculations 
are the photon two-point function at $q^2=0$ and the $\gamma-Z^0$ mixing 
amplitude at $q^2=\mz^2$. They 
are the only contributions to   $\delta^{(2)} e/e|_{\msbar}$
and $\Delta\hat{k}^{(2)}$, respectively, and therefore can be read
directly from \eqs{deltae} and (\ref{deltak2}).  

The limit for $q^2\to 0$ of the previous expressions can be easily obtained 
using the expansion of $B_0[q^2,a,b]$
in powers of $q^2$:
\bea
B_0[q^2,a,b]=   1- 
   {\frac{a\,\ln a - b\,\ln b}{a - b}}
+ q^2 {\frac{\left( a^2 - b^2 - 
         2\,a\,b\,\ln {\frac{a}{b}} \right) }{{{2\left( a - b \right) }^3}}} 
+\,O(q^4)
\eea

As an example of the use of the above self-energies, we now calculate the 
$\msbar$ counterterm $\delta\scur$ at \gd. We recall that, following 
 Eq.\,(12b) of \efe{dfs}, $\delta\scur=-\mw^2/\mz^2 \, \, Y|_{pole}$, 
where $Y$ is the   expression 
appearing on the r.h.s of \equ{quattro}, before the poles are subtracted. 
Inserting the poles of
\eqs{azzlh} and (\ref{awwlh}) or \eqs{azzhh} and (\ref{awwhh}) into
the previous relation is not
sufficient to obtain the correct result in the pure $\msbar$ scheme because
(i) no one-loop coupling renormalization has yet been carried out and (ii) 
the two-loop self-energies 
have been obtained using OS masses instead of a pure $\msbar$
subtraction. The $\msbar$ renormalization of gauge couplings does not introduce
$\mt^2 $ terms and at \amtd it is only needed  for  the leading quadratic
contribution. Working as usual in $n$ dimensions, we find 
\bea
-\frac{\delta \hat{g}^2}{\hat{g}^2} Y_{\msbar}^{(1, lead)}=
-\frac{19}{6} \frac{N_c\,\hat{g}^4}{(16\pi^2)^2} \frac{\mt^2}{4\mw^2}
\left(\frac1{\eps} +\frac1{2} -\ln \frac{\mt^2}{\mu^2}+O(\eps)\right).
\label{renY1}
\eea
In order to re-express the poles of the
two-loop self-energies in terms of $\msbar$ quantities we take \eqs{azz1} and
(\ref{aww1}), insert them in $\mw^2/\mz^2 \ Y$, 
replace $\mw^2/\mz^2$ by $\hat{c}^2
(1+\frac{\hat{e}^2}{\hat{s}^2}
\drhoc^{(1)}_{lead})$ in $n$ dimensions, and expand at \amtd.
The extra term at this order reads
\be
Y^{(2)}_{extra}=
-             \frac{N_c\,\hat{g}^4}{(16\pi^2)^2} \frac{\mt^2}{4\mw^2}
\left(\frac1{\eps} +\frac1{2} -\ln \frac{\mt^2}{\mu^2} +O(\eps)       \right)
\left(4\ccur - \frac{13}{6} +\frac{1}{\ccur}\right) 
\label{Y2add}.
\ee
Combining the poles of the two-loop self-energies 
in $Y$ together with \eqs{renY1}
and  $Y^{(2)}_{extra}$, we find the correct $\msbar$ counterterm
\be
\frac{\delta^{(2)}\scur}{\scur}=
-\frac{N_c\, \gc^4}{(16\pi^2)^2}\, \frac{\mt^2}{\ccur\mz^2}\,\frac1{\epsilon}
\left(\frac1{8} - \frac{13}{36} \,\scur\right),
\ee
which can be checked with \cite{macha}. Combining the finite parts of these
three components, instead, one obtains  $\drhoc^{(2)}$ given in \cite{dgv}.

\end{appendletterA}
  \setcounter{section}{1}
\begin{appendletterB}
\section*{Appendix B}
\label{appB}
\appendix
In this Appendix we present the two-loop  \gd\ contributions to the
electroweak corrections entering  the $Z$ decay partial widths.

We begin by considering the $\hat{\eta}$ and $\hat{k}$ corrections of the
$\msbar$ framework. We have for $\mh \ll \mt$ in units 
$N_c/(16 \pi^2) \,(\hat{\alpha}/4\pi\scur) \, \mt^2/(\mz^2 \ccur) $
\bea\Delta\hat{\eta}^{(2)}
&=&
  {{{h_t^3} - 6\,{h_t^2}\,{z_t} + 
      11\,{h_t}\,{{{z_t}}^2}}\over 
    {9\,{\ccur}\,\left( {h_t} - 4\,{z_t} \right) \,{z_t^2}}}
+  {{49 - 289\,\ccur - 349\,\hat{c}^4 + 292\,\hat{c}^6}\over 
    {216\,\ccur\,\left( 1 - 4\,\ccur \right) }}
\nn\\& +&
 {{1 + 18\,\ccur - 16\,\hat{c}^4}\over {12\,\left( 1 - 4\,\ccur \right) }}
\ln \ccur -
{{17 - 40\,\ccur + 32\,\hat{c}^4}\over {54\,\ccur}}
({\sqrt{{ h_t}}}\,\pi -2)
\nn\\&+&
{{11\,{ h_t^2}\,{ z_t}-2\,{ h_t^3}  - 
      24\,{ h_t}\,{ z_t^2} + 24\, z_t^3}\over 
    {18\,\ccur\,\left( { h_t} - 4\,{ z_t} \right) \,{ z_t^2}}}\ln h_t +
  {{1 - 4\,\ccur + 44\,\hat{c}^4 -32\,\hat{c}^6}\over {24\,\ccur (1-4\,c^2)}}
B_0[z,w,w] 
\nn\\&+&
  {{ 13\,{ h_t^2}\,{ z_t}  -2\,{ h_t^3} -
      32\, h_t\,{ z_t^2} + 36\, z_t^3}\over 
    {18\,\ccur\,\left( { h_t} - 4\,{ z_t} \right) \,{{{ z_t}}^2}}}
B_0[z,h,z] 
-{{17 - 34\,\ccur + 26\,\hat{c}^4}\over {36\,\ccur}}
\ln \frac{\mt^2}{\mu^2} 
\nn\\&+&
\left(
   {{{ h_t\,(2h_t-5z_t)}}\over {
18\,\ccur\,z_t\,\left( { h_t} - 4\,{ z_t} \right) }} + 
  {{10 - 39\,\ccur - 70\,\hat{c}^4 + 48\,\hat{c}^6}\over 
    {36\,\ccur\,\left( 4\,\ccur-1 \right) }}\right)\ln z_t.
\label{eta2}
\eea
In the case of $\mh\gg \mz$ we obtain in the same units
\bea
\Delta\hat{\eta}^{(2)}&=&
  {{\left( -17 + 40\,\ccur - 32\,\hat{c}^4 \right) 
{h_t}}\over {216\,\ccur}}+
{5\over {144\,\ccur\,\left({h_t} -4\right) }}
+  {{707 - 4720\,\ccur + 5900\,\hat{c}^4 - 3696\,\hat{c}^6}\over 
    {864\,\ccur\,\left( 1 - 4\,\ccur \right) }}
\nn\\&+&
\left(  {{10}\over {27}} - {{17}\over {108\,\ccur}} - {{8\,\ccur}\over {27}}
\right)(1-\frac{h_t}{4}) \sqrt{h_t}\, g(h_t)+  
{{1 + 18\ccur - 16\hat{c}^4}\over {12\,\left( 1 - 4\,\ccur \right) }}
\ln \ccur +
{{4 - {h_t}}\over {12\,\ccur\,{h_t}}} \Lambda(h_t)
\nn\\&+&
  {{2 - 7\,\ccur - 70\,\hat{c}^4 + 48\,\hat{c}^6}\over 
    {36\,\ccur\,\left(  4\,\ccur -1\right) }} \ln z_t +
  {{1 - 4\,\ccur + 44\,\hat{c}^4 - 32\,\hat{c}^6}\over 
    {24\,\ccur\,\left( 1 - 4\,\ccur \right) }}
B_0[z,w,w] 
\nn\\&-&
{{17 - 34\ccur + 26\hat{c}^4}\over {36\,\ccur}}
\ln \frac{\mt^2}{\mu^2} 
+
\left[
{{(4\ccur -5) 
      \left( 6 + 27{h_t} - 10{ h_t^2} + {h_t^3} \right) }
\over {54\left( {h_t}-4 \right) }}\right.
\nn\\&-&\left.
  {{1152 + 606{h_t} + 1467{{{h_t}}^2} - 1097{{{h_t}}^3} + 
      238{{{h_t}}^4} - 17{{{h_t}}^5}}\over 
    {432\,\ccur{{\left( {h_t}-4 \right) }^2}\,{h_t}}}
\right]\ln h_t
\nn\\&+&
\left[
  {{\left( 5 - 4\,\ccur \right) \,\left( {h_t}-1 \right) }\over 
    {9\,\left( {h_t} -4\right) \,{h_t}}}
-  {{384 + 10\,{h_t} - 238\,{{{h_t}}^2} + 63\,{{{h_t}}^3} - 
      3\,{{{h_t}}^4}}\over 
    {144\,\ccur\,{{\left( {h_t}-4 \right) }^2}\,{{{h_t}}^2}}}
\right]\phi(\frac{h_t}{4}).
\label{eta2b}
\eea

The correction $\hat k$ can be written as 
$\hat{k}_f = 1 + (\ecur^2 /\scur ) (\Delta\hat{k}_f^{(1)} +
\Delta\hat{k}_f^{(2)})$ where the two-loop part has been given
already in \efe{dgs}. We report here again for completeness.
Considering $\mh \gg \mt$ 
in units $N_c/(16 \pi^2) (\hat{\alpha}/(4\pi \hat{s}^2)) \mt^2/(c^2 \mz^2)$ 
one has
\begin{eqnarray}
{\rm Re}\,\Delta\hat{k}^{(2)}(\mz^2)&=&
{{-211 + 24\,{h_t} + 462\,{\scur} - 64\,{h_t}\,{\scur
}}\over {432}} + 
  \left( {3\over 8} - {{{\scur}}\over 3} \right) 
   B_0[z,w,w]  -{\ccur\over6} \ln \ccur 
\nonumber\\&+&
  {{\left( h_t-4 \right) \,{\sqrt{{h_t}}}\,
      \left(  8\,{\scur}-3 \right) g(h_t)}\over {108}} - 
  {{\left( 6 + 27\,{h_t} - 10\,{{{h_t}}^2} + {{{h_t}}^3} \right) \,
      \left( 3 - 8\,{\scur} \right) }\over 
    {108\left( {h_t}-4 \right) }} \ln {h_t}
\nonumber\\&-&
  \left( {1\over4} + {2\over9}{\scur} \right) \ln {\mt^2\over\mu^2}+  {{\left(  3\,{\scur} -2\right) }\over {18}} \ln z_t + 
  {{\left(  {h_t}-1 \right) \left(  8\,{\scur} -3\right) 
     }\over 
    {18\left( 4 - {h_t} \right) {h_t}}} \, \phi ({h_t\over 4}).
\label{deltak2}
\end{eqnarray}
In the case of a light Higgs mass, this becomes
\bea
{\rm Re}\,\Delta\hat{k}^{(2)}(\mz^2)&=&
{\frac{-175 + 366\,\scur}{432}}
+  \left( {3\over 8} - {{{\scur}}\over 3} \right) 
   B_0[z,w,w]  -{\ccur\over6} \ln \ccur 
\nonumber\\&-&
\frac{2\pi}{27} \sqrt{h_t} \left(8\scur -3 \right)
-  \left( {1\over4} + {2\over9}{\scur} \right) \ln {\mt^2\over\mu^2}+  {{\left(  3\,{\scur} -2\right) }\over {18}} \ln z_t 
\end{eqnarray}
Numerically, 
 ${\rm Re}\,\Delta\hat{k}^{(2)}(\mz^2)$
is  tiny, and the use of the  expansion for light 
Higgs   does not lead to any appreciable difference over the other one.

We give also the two-loop finite part of the electric charge 
counterterm, which is needed to calculate the $\msbar$ electromagnetic
coupling. We have  in units $N_c
  \,\hat{\alpha}^2/(16\pi^2 s^2) \mt^2/(\mz^2 \ccur) $
\bea
\left.\frac{2\delta^{(2)}e}{e}\right|_{\msbar}&=& 
  {{231 - 32\,{h_t}}\over {216}} - 
  \frac{2}{27} ( 4 - {h_t} ) \,{\sqrt{{h_t}}}\,g(h_t) + 
   {{2\left( 6 + 27\,{h_t} -10\,{{{ h_t}}^2} + {{{h_t}}^3} \right)
        }\over {27\,\left({h_t}-4 \right) }} \ln h_t 
\nn\\&-& 
    \frac{13}{18}  \ln {\mt^2\over\mu^2}
 -  {{4\,\left(  {h_t}-1 \right) \,\phi ({{{h_t}}\over 4})}\over 
     {9\,\left( {h_t}-4 \right) \,{h_t}}},
\label{deltae}
\eea
 that in the case of a light Higgs  reduces to 
$  {{61}\over {72}} - {{16\,{\sqrt{{h_t}}}\,\pi }\over {27}}  
   -\frac{13}{18}  \ln {\mt^2\over\mu^2}$.

We proceed to the discussion of the corrections of the OS scheme.
Recalling \equ{etadef2} the additional two-loop contribution in
$\Delta \bar{\eta}_f^{(2)}$ is given by (units $N_c x_t/(16\pi^2)$:
\bea
\Delta\bar\eta^{(2)}_{f,add}&=& 16\pi^2 \Delta\bar\eta^{(1)}_f
+ V_{add}
-  {{197 - 1378\,{c^2} + 1064\,{c^4}}\over 
     {27\,\left( 1 - 4\,{c^2} \right) }} \nonumber\\ 
&&- 
   {{\left( 1 + 16\,{c^2} - 20\,{c^4} + 48\,{c^6} \right) \,
 \over     {3 (1-4c^2)}}        B_0[z,w,w]}
 - {{2\,{c^2}\,
       \left( 1 + 26\,{c^2} + 24\,{c^4} \right) }\over 
     {3 (1-4c^2)}} \ln c^2\nonumber\\
&&+\left({{41}\over 3} - {{46\,{c^2}}\over 3} \right) 
\ln \frac{\mt^2}{\mu^2} + 
   {{2\,\left( 50 - 283\,{c^2} + 242\,{c^4} - 72\,{c^6} \right) 
      }\over {9 (1-4c^2) }} \ln  z_t.
\label{etadd}
\eea
where $V_{add}$ contains the shifts of the one-loop vertices.
\bea
V_{add}=8 c^2 \ln \frac{\mw^2}{\mu^2} + 
3 \left(I^3_f Q_f - 4s^2 Q_f^2\right) f_\smallv(1) - 
16c^2 g_\smallv(c^{-2})\nonumber\\
+\left(1-4c^2 -2(1-2c^2) I^3_f Q_f\right) f_\smallv(c^{-2})
\label{vadd}
\eea
and the functions $f_V(x)$ and $g_V(x)$ are defined in Eqs.(6d) and (6e)
of \cite{ds91}. 

The form factor $k_f$ was considered in \efe{dgs} only for the leptonic case.
The generalization of Eq.~(A4) of that reference to arbitrary fermions is
(units  $N_c x_t/(16\pi^2)$)
\bea
\Delta \bar k_{add,f}^{(2)}=&& {{-238\,{c^2}}\over {27}} + 8{c^4} - 
   2\,{c^2}\,{\sqrt{ 4 c^2 -1}}\left( 3 + 4\,{c^2} \right) 
    \arctan ({1\over {{\sqrt{ 4\,{c^2} -1}}}}) - 
   {16\over 9}c^2\ln z_t
\nonumber\\&&
+  {\frac{1 - 6\,I_f^3\,Q_f + 8\,Q_f^2 - 
      8\,{c^4}\,Q_f^2}{4\,{c^2}}}  f_V(1) + 
   4\,{c^2}{ g_V}({c^{-2}}) - 7\,{c^2}\,\ln c^2 - 
   {{17\over3}}c^2 \ln {\mu^2\over \mz^2} 
\nonumber\\&&
+ c^2 (1-Q_f I^3_f) f_V(c^{-2})
-\frac{80}{9} i \pi. 
\label{dkaddf}
\eea

To complete the discussion in the OS framework we need also the
corrections $\Delta\bar\rho (s^2)$ and $\bar{f}(s^2)$. 
They have been considered in \efe{dgs}. Recalling the definitions
\bea
\bar{f}(s^2)&=& \hat{f}^{(1)}(s^2) + 
\bar{f}^{(2)}(s^2) 
\label{dodici}\\
\Delta\bar{\rho}(s^2)&=& \Delta\hat{\rho}^{(1)}(s^2) + \Delta\bar{\rho}^{(2)}
(s^2)
\label{tredici},
\eea
where
 $\bar{f}^{(2)}(s^2)= \hat{f}^{(2)}(s^2)+ \bar{f}^{(2)}_{add}(s^2)$,
$\Delta\bar{\rho}^{(2)}(s^2)=\Delta\hat{\rho}^{(2)}(s^2)+
\Delta\bar{\rho}^{(2)}_{add}(s^2)$, we have that
$(G_\mu/\sqrt{2})\, 8 \mw^2 \, \Delta\hat{\rho}^{(2)}$ is given 
by Eqs.(10a,b) of \efe{dgv} multiplied by $N_c x_t^2$ while
$\Delta\bar{\rho}^{(2)}_{add}(s^2)$ is reported in Eq.(A2) of \efe{dgs}.
The amplitude $(e^2/s^2) \hat{f}^{(2)}$ is given by the  sum of
$(2\delta^{(2)}e/e)$ given in \equ{deltae}
and  Eqs.(7a,b)\footnote{Eq.(7b) of 
\efe{dgv} has a misprint in the fourth term
of the first line. The correct term is $77 c^2/72$ instead of $77 c^2/12$
as printed in \efe{dgv}. We thank G.~Passarino for pointing this out to us.}
of \efe{dgv} multiplied by $(\alpha/\pi s^2) N_c\, x_t$.
The term $ \bar{f}^{(2)}_{add}(s^2)$ is given in Eq.(A1) of \efe{dgs}.

\end{appendletterB}

\end{document}